\documentclass[suppldata]{interact}

\usepackage{epstopdf}
\usepackage[caption=false]{subfig}
\usepackage{enumitem}
\usepackage{pdfpages}

\usepackage[natbibapa,nodoi]{apacite}
\newcommand{\tnote}{\, (Table 1)}

\usepackage{color}
\usepackage{graphicx}
\usepackage{rotate}
\usepackage{lineno}
\usepackage{lscape}

\theoremstyle{plain}

\theoremstyle{definition}

\theoremstyle{remark}

\begin{document}

\title{New Science, New Media: An Assessment of the Online Education and Public Outreach Initiatives of The Dark Energy Survey}

\author{\name{R.~C.~Wolf\textsuperscript{1,2}\thanks{CONTACT: R.~C.~Wolf; rcwolf@stanford.edu}, A.~K.~Romer\textsuperscript{3}
, {B.~Nord}\textsuperscript{4}, {S.~Avila}\textsuperscript{5}, {K.~Bechtol}\textsuperscript{6}, {L.~Biron}\textsuperscript{4}, {R.~Cawthon}\textsuperscript{7}, {C.~Chang}\textsuperscript{7}, {R.~Das}\textsuperscript{8}, {A.~Fert\'e}\textsuperscript{9,10}, {M.~S.~S.Gill}\textsuperscript{11},
{R.~R.~Gupta}\textsuperscript{12},
{S.~Hamilton}\textsuperscript{8}, {J.~M.~Hislop}\textsuperscript{3}, {E.~Jennings}\textsuperscript{13}, {C.~Krawiec}\textsuperscript{2},
{A.~Kremin}\textsuperscript{8},
{T.~S.~Li}\textsuperscript{4},
{T.~Lingard}\textsuperscript{5,3},
{A.~M{\"o}ller}\textsuperscript{14},
{J.~Muir}\textsuperscript{8},
{D.~Q.~Nagasawa}\textsuperscript{15}, {R.~L.~C.~Ogando}\textsuperscript{16,17}, {A.~A.~Plazas}\textsuperscript{18},
{I.~Sevilla-Noarbe}\textsuperscript{19},
{E.~Suchyta}\textsuperscript{20},
{Y. Zhang}\textsuperscript{4},
{J. Zuntz}\textsuperscript{9}
}
\affil{\textsuperscript{1}Graduate School of Education, Stanford University, 160, 450 Serra Mall, Stanford, CA 94305, USA; \textsuperscript{2}Department of Physics and Astronomy, University of Pennsylvania, 209 South 33rd Street, Philadelphia, PA 19104, USA; \textsuperscript{3}Department of Physics and Astronomy, Pevensey Building, University of Sussex, Brighton, BN1 9QH, UK; \textsuperscript{4}Fermi National Accelerator Laboratory, P. O. Box 500, Batavia, IL 60510, USA; \textsuperscript{5}Institute of Cosmology \& Gravitation, University of Portsmouth, Portsmouth, PO1 3FX, UK; \textsuperscript{6}LSST, 933 North Cherry Avenue, Tucson, AZ 85721, USA; \textsuperscript{7}Kavli Institute for Cosmological Physics, University of Chicago, Chicago, IL 60637, USA; \textsuperscript{8}Department of Physics, University of Michigan, Ann Arbor, MI 48109, USA; \textsuperscript{9}Institute for Astronomy, University of Edinburgh, Blackford Hill, Edinburgh EH9 3HJ, UK; \textsuperscript{10}Department of Physics and Astronomy, University College London, Gower Street, London WC1E 6BT, UK; \textsuperscript{11}Kavli Institute for Particle Astrophysics and Cosmology, Department of Physics and SLAC National
Accelerator Laboratory, Stanford University, Stanford, CA 94305, USA; \textsuperscript{12}Lawrence Berkeley National Laboratory, 1 Cyclotron Road, Berkeley, CA 94720, USA; \textsuperscript{13}Argonne National Laboratory, Leadership Computing Facility, Lemont, IL, 60439, USA; \textsuperscript{14}Research School of Astronomy and Astrophysics, Australian National University, Canberra, ACT 2611, Australia
ARC Centre of Excellence for All-sky Astrophysics (CAASTRO); \textsuperscript{15}George P. and Cynthia Woods Mitchell Institute for Fundamental Physics and Astronomy, and Department of Physics and Astronomy, Texas A\&M University, College Station, TX 77843, USA; \textsuperscript{16}Observat\'orio Nacional, Rua Gal. Jos\'e Cristino 77, Rio de Janeiro, RJ - 20921-400, Brazil; \textsuperscript{17}Laborat\'orio Interinstitucional de e-Astronomia - LIneA, Rua Gal. Jos\'e Cristino 77, Rio de Janeiro, RJ - 20921-400, Brazil; \textsuperscript{18}Jet Propulsion Laboratory, California Institute of Technology, 4800 Oak Grove Dr., Pasadena, CA 91109, USA; \textsuperscript{19}Centro de Investigaciones Energ\'eticas, Medioambientales y Tecnol\'ogicas (CIEMAT), Madrid, Spain; \textsuperscript{20}Computer Science and Mathematics Division, Oak Ridge National Laboratory, Oak Ridge, TN 37831, USA} 
}

\maketitle

\begin{abstract}
As large-scale international collaborations become the standard for astronomy research, a wealth of opportunities have emerged to create innovative education and public outreach (EPO) programming. In the past two decades, large astronomy collaborations, such as Hubble and the Sloan Digital Sky Survey, have primarily focused their EPO strategies around published data products. Newer collaborations, however, have begun to explore other avenues of public engagement both before and after the data are made available. We present a case study of the online EPO strategy and products of The Dark Energy Survey, currently one of the largest international astronomy collaborations actively taking data. The DES EPO program is unique at this scale in astronomy, as far as we are aware, as it evolved organically from scientists' passion for EPO and is entirely organized and implemented by the volunteer efforts of collaboration scientists. We summarize the strategy and implementation of eight informal astronomy education and engagement programs and present evidence-based recommendations for future projects. For content distributed via social media, we present reach and user statistics over the 2016 calendar year.  Using native social media platform analytics, we find that DES EPO online products reached an average of 2,500 users per post. We also find that 94\% of these users indicate a predisposition to science-related interests. We find no obvious correlation between post type and post reach, with the most popular posts featuring the intersections of science and art and/or science and popular culture. Upon evaluation, we find that one of the key issues of the online DES EPO program was designing material which would inspire new interest in science. The greatest difficulty of the online DES EPO program was sustaining scientist participation and collaboration support; we find the most successful programs in this regard are those which capitalized on the hobbies of the participating scientists.  We present these statistics and recommendations, along with observations from individual experience, as a potentially instructive resource for other scientists or EPO professionals interested in organizing EPO programs and partnerships for large science collaborations or organizations.
\end{abstract}


\section{Introduction}
\label{S:1}
The landscape of professional astronomy has dramatically transformed over the past fifty years. While the field was once dominated by individuals or small, co-located teams (e.g., a professor and a graduate student), advances in technology have revolutionized the ways in which science is practiced and communicated \citep{NewHor}. International collaborations have emerged as the new standard. The primary charge of these large-scale astronomy projects is to use evidence-based research to answer fundamental questions about our Universe. The drive to solve mysteries such as the nature of dark matter and dark energy drive project commissioning, instrument development, project implementation, data products, and analysis \citep{NewHor}. Collaboration on such a large scale requires cooperation and respect amongst scientists from a diverse group of ages, genders, and cultures. In addition to the potential for groundbreaking science, this next generation of astronomy collaborations also comes with a wealth of innovative material and experience that can be used to capture the public's interest in science \citep{Borne10}.

Education and Public Outreach (EPO) programming has been a cornerstone of nationally sponsored agencies such as the National Aeronautics and Space Administration (NASA) in the USA for decades. The NASA Office of Education is dedicated to designing hands-on activities, creating teacher resources, developing opportunities for students, and inspiring students to pursue careers in the science, technology, engineering, and math (STEM) disciplines \citep{Rosendhal2004}. Similarly, the multi-national European Space Agency (ESA) has a well-developed and actively maintained EPO presence.\footnote{\texttt{http://www.esa.int/Education}} The NASA and ESA EPO efforts would not be possible without an agreed strategy, support from dedicated, well-trained staff, and an appropriate funding stream.

In the past ten years, many large-scale astronomy programs have devoted funding and personnel to EPO programming, for both in-person activities (e.g., materials for scientists visiting K-12 classrooms and attending science festivals) and online projects. For example, the Hubble Space Telescope (HST) and Sloan Digital Sky Survey created their own EPO initiatives \citep{Griffin2003,Raddick2002}, including Hubblesite\footnote{\texttt{http://hubblesite.org/}} and SDSS Voyages,\footnote{\texttt{http://voyages.sdss.org/}} which encourage web-based users to explore publicly available astronomy images and data products through a variety of online lesson plans and hands-on activities. In addition to more conventional avenues of astronomy EPO such as public lectures, science festivals, and planetarium shows, several innovative avenues for connecting science, and scientists, with the public have emerged. For example, {\it citizen science}, whereby expert scientists collaborate with members of the public to complete a science project, is growing in popularity year-after-year \citep{Borne10,Haywood14}. 

The importance of EPO activities to modern astronomy is demonstrated by the fact that several projects that are still in the development stage are already investing resources into public engagement. For example, the website of the James Webb Space Telescope (JWST, set to launch in 2019), already includes detailed EPO materials designed for K-12 formal and informal education.\footnote{\texttt{https://jwst.nasa.gov/teachers.html}} The Large Synoptic Survey Telescope (LSST), which will not begin taking data until the end of the decade, describes its EPO program as ``as ambitious as the telescope itself.''\footnote{\texttt{https://www.lsst.org/about/epo}} The LSST EPO program includes plans for citizen science partnerships with The Zooniverse,\footnote{\texttt{https://www.zooniverse.org/}} and data visualization projects with several planetaria. Finally, The Wide-Field Infrared Survey Telescope (WFIRST, set to launch in the mid 2020's) outlines the internal organizational structure of the project and includes EPO as an element of its Science Operations Center.\footnote{\texttt{https://wfirst.gsfc.nasa.gov/science/}}

As the examples above demonstrate, EPO programming is now being put at the forefront of collaboration structure well before any data have been taken. In the case of the Dark Energy Survey (DES, founded in 2004), EPO was not embedded during the development stage and had to be ``shoehorned in'' after survey operations were underway. The full DES EPO program, which includes both online initiatives and in-person lesson plans and demonstrations, evolved from the grass-roots efforts of a small number of collaboration members who are passionate about science communication and outreach. This ``bottom-up'' approach has been positive in that it has resulted in a variety of innovative EPO projects. However, there have also been some unforeseen challenges. As such, our EPO experience in DES provides a unique perspective that can be used to inspire (and/or caution) teams of scientists and EPO professionals developing EPO programs for the next generation of large astronomy projects. 

Throughout this work, we present a detailed case study of the \textbf{online} DES EPO initiatives which were part of the DES EPO program during its first three years. We remind the reader that this text does not describe all DES EPO projects; we have focused this analysis on online initiatives as an introduction to the DES EPO repertoire and as such projects have quantitative metrics which can be used as an example measure of reach. Several discussions include how various resources were allocated for EPO activities. We note that throughout this article, ``resources'' refers to both the time spent by DES member volunteers on EPO activities and to the financial support provided, at the discretion of the DES Director, from participating institutions. In Section~\ref{sec:desover} we outline the guiding principles which motivated these projects. In Section~\ref{sec:programs}, we describe several of the online DES EPO initiatives, including project goals, organization details, and project outputs. For each, we propose recommendations for similar future large-scale collaborations. Relevant DES project links are listed in Table~\ref{onlineres}. We conclude with a summary of the DES EPO experience thus far in Section~\ref{sec:conclusion}. For a description of DES science goals and organizational structure, see Appendix~\ref{sec:sciproj}.

\begin{table}[htb]
\centering
\begin{tabular}{l l}
\hline \\[-6.5pt]
EPO Product & Website URL \\[5.5pt]
\hline \\[-6.5pt] 
DES Website & \footnotesize{\texttt{www.darkenergysurvey.org}}  \\[5.5pt]
DES Facebook & \footnotesize{\texttt{www.facebook.com/darkenergysurvey}} \\[5.5pt] 
DES Twitter & \footnotesize{\texttt{www.twitter.com/theDESurvey}} \\[5.5pt] 
Dark Energy Detectives & \footnotesize{\texttt{www.darkenergydetectives.org}}\\[5.5pt] 
The DArchive & \footnotesize{\texttt{www.darkenergysurvey.org/news-and-results/darchives/}} \\[5.5pt] 
DarkBites & \footnotesize{\texttt{www.darkenergysurvey.org/education/darkbites/}} \\[5.5pt]
DEScientist of the Week & \footnotesize{\texttt{www.darkenergysurvey.org/education/scientist-of-the-week/}}\\[5.5pt]
DES Flickr & \footnotesize{\texttt{www.flickr.com/photos/129954880@N03}}\\[5.5pt]
DES YouTube Channel & \footnotesize{\texttt{www.youtube.com/channel/UCkAD7Un4aX--Y2ETTs\_mImQ}} \\[5.5pt]
\hline \\[1.5pt]
\end{tabular}
\caption{List of Online DES EPO Products.} \label{onlineres}
\end{table}

Below, we outline the most significant lessons learned from the DES EPO experience and encourage the reader to see relevant sections throughout the article for further discussion.

\begin{itemize}
\item{Organizing EPO for a large collaboration has significant advantages and challenges. An EPO program which relies on the participation of collaborating scientists should be embedded into collaboration infrastructure and culture \textit{from the outset}.}
\item{The most sustainable projects are those which capitalize on the collective creativity of scientists.}
\item{Social media is a useful tool to engage scientists in EPO and communicate science to the general public. In 2016, the average DES EPO social media post reached $\sim$$2,500$ users. The reach of social media products both within and between projects was highly variable, with the lowest- and highest-reaching posts reaching 258 and 14,450 users, respectively.}
\item{Despite the logistical advantages of social media programming (e.g., the ability to engage a large diverse audience, or the capacity for scientists to engage in EPO from anywhere in the world), we do not have evidence that DES EPO products inspire new interest in science. According to native platform analytics, 94\% of reached social media users demonstrate a predisposition to science-related topics and interests. This suggests that EPO organizing committees should place particularly careful thought into the intended audience and goals of online programming.}
\item{It is difficult to incentivize scientists to participate in EPO programs unless they are already inclined to do so.}
\end{itemize}

\section{Guiding Principles and Social Media Strategy} 
\label{sec:desover}

We begin this section with an outline of the primary guiding principles which motivated the DES EPO effort (Section~\ref{subsec:hypo}). These principles were formulated from the previous experiences of the founding co-coordinators of the DES Education \& Public Outreach Committee (EPOC): B.~Nord, A.~K.~Romer, and R.~C.~Wolf (hereafter NRW; the three primary authors of this article). 

Over time, the EPOC grew to include other collaboration members. Further details regarding DES science, organizational structure, and the evolution of the EPOC are described in Appendix~\ref{sec:sciproj}. We conclude this section with a summary of the DES EPO social media strategy as social media has been a primary vehicle for DES EPO online product distribution (Section~\ref{subsec:sm}).


\subsection{DES EPO Guiding Principles}
\label{subsec:hypo}

\begin{itemize}
\item{GP1: EPO is an important, worthwhile, and enjoyable activity for individual scientists.}
\item{GP2: EPO is an important and worthwhile activity for science collaborations (especially those which benefit from public funding).}
\item{GP3: The public are interested in scientists as well as the science.}
\item{GP4: It is possible to challenge the public's perception of scientists (as ``old white males'') through EPO.}
\item{GP5: DES should have a strong social media presence. This should consist of frequent posts with high-quality content.} 
\item{GP6: DES should have a professional website with high-quality embedded content.}
\item{GP7: DES EPO should not be restricted to the English language.}
\item{GP8: The broader DES membership would be motivated by the EPOC when presented with EPO opportunities.}
\item {GP9: The EPOC will manage, organize, and coordinate all DES EPO efforts.}

\end{itemize}

These tenets laid the foundation for nearly three years of DES EPO programming. They also informed interactions between the EPOC and other groups within the DES organizational structure (see Appendix~\ref{subsec:org}). While the DES EPO program at large was motivated by GP1 and GP2, specific initiatives had more targeted goals. Implemented DES EPO initiatives were drawn from a substantial list of creative ideas, but were limited by the lack of time that the EPOC could contribute in addition to their regular duties (and to a lesser extent, the lack of a dedicated funding stream). In Section~\ref{sec:programs}, we detail specific projects and how, where relevant, these principles contributed to project development. We stress that these guiding principles stemmed from the previous EPO experience of NRW, rather than being informed by the literature in the science communication field. In hindsight, it is clear that the EPOC would have benefited from some external guidance before launching into project work (see Section~\ref{sec:conclusion}).

\subsubsection{Diversity and Inclusivity}
\label{subsubec:DnI}

NRW followed what they believed were best practices in developing materials for dissemination on social media (based on private conversations with communication professionals), and in convening and organizing the outreach efforts of colleagues. In particular, addressing the lack of diversity and inclusion in scientific work environments \citep{Ivie2014, Stassun} was a consistent thread in our activities, despite not being an explicit Guiding Principle.  For example, NRW put thought into project design such that all language and visual information was presented in way so as to be inclusive of under-represented minorities (of all genders) and white women. Additionally, NRW recruited DES scientists from diverse backgrounds to participate in EPO activities (both online and in person). We urge other large collaborations/organizations planning EPO science programs to include diversity and inclusivity as fundamental organizing pillars.

\subsection{DES Social Media Strategy and User Summary}
\label{subsec:sm}

Social media (e.g., websites and applications used for content sharing and social networking) has been one of the main vehicles for distribution of DES EPO content online. Facebook and Twitter are two of the many popular social media platforms currently available, and were the platforms of choice for DES EPO. The DES Facebook account was created in November 2010; the Twitter account followed in October 2013\tnote. Prior to the creation of the EPOC, social media posts were sparse, with updates roughly once per month. When the EPOC formed in October 2014, NRW decided to centralize the social media effort and make regular posts a priority (GP3, GP5, GP9; Section~\ref{subsec:hypo}). For this reason, most of the online DES EPO projects have been driven by the need for a regular stream of social media content. Since early 2015, there have typically been Facebook (and mirrored Twitter) posts at least five times per week: DEST4TD (Section~\ref{subsec:t4td}) on Mondays, Tuesdays, and Thursdays, DarkBites (Section~\ref{subsec:darkbites}) on Wednesdays, DEScientist of the Week (Section~\ref{subsec:sow}) on Fridays, and MLDES (Section~\ref{subsec:transla}) on Sundays. We note that we opted to post content manually per platform, rather than use a social media management dashboard.  


From October 2014 to April 2017, the number of Facebook page ``likes'' increased from $\sim$$ 5,100$ to $\sim$$8,000$ and the number of Twitter followers increased from $\sim$$30$ to $\sim$$1,400$. Since the formation of the EPOC, both the DES Facebook and Twitter followings have increased roughly linearly, with $\sim$$5$ new followers added per day.


Throughout this article, we present various
metrics used to assess the social media strategy and reach of several DES EPO initiatives.\footnote{These metrics were taken at the time of writing: April 2017.} We use these metrics as an initial measure of impact, but note that they do not provide a complete picture of how users interact with, and learn from, DES EPO material. Many of these metrics were obtained from Facebook Insights and Twitter Analytics (native, integrated services offered by the respective social media platforms to the user for account analysis). Any categorical information (e.g., the gender of the users) used from these metrics is defined by the platforms. We have also conducted formative surveys within the collaboration (e.g., Appendix~\ref{sec:intersur}) and extracted our own data from the social media sites.

\begin{figure}[htp]
\includegraphics[scale=0.35]{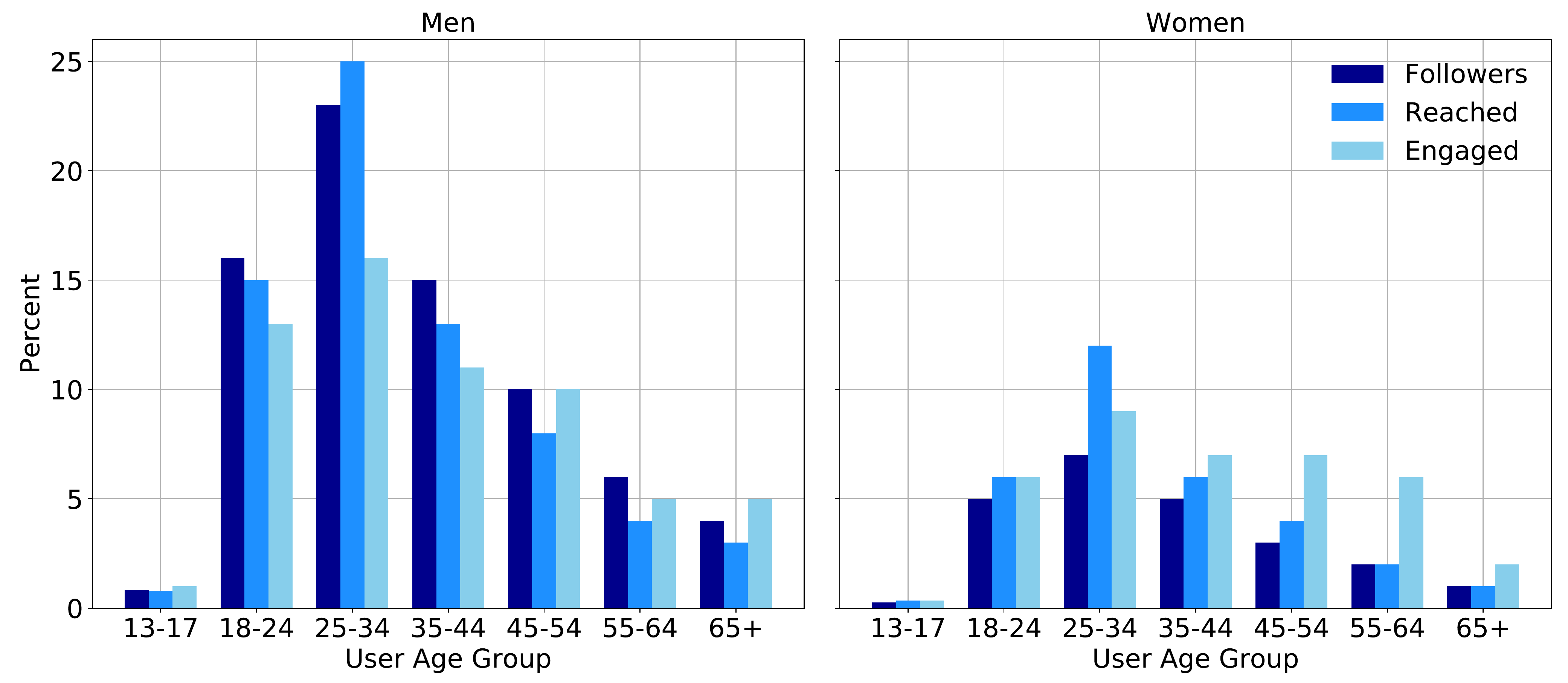}
\caption{Percent of Facebook followers (dark blue), users reached (blue), and engaged users (light blue) over various user ages. Information is also separated into men (left) and women (right) user groups. The 25-34 user age group is the most popular in each user group and for both men and women. While the men are roughly 75\% of the total page followers $(n=7914)$, women make up nearly 40\% of the total engaged users $(n\approx 500)$.}
\label{fig:menwom}
\end{figure}

\begin{figure}[htp]
\centering
\includegraphics[scale=0.35]{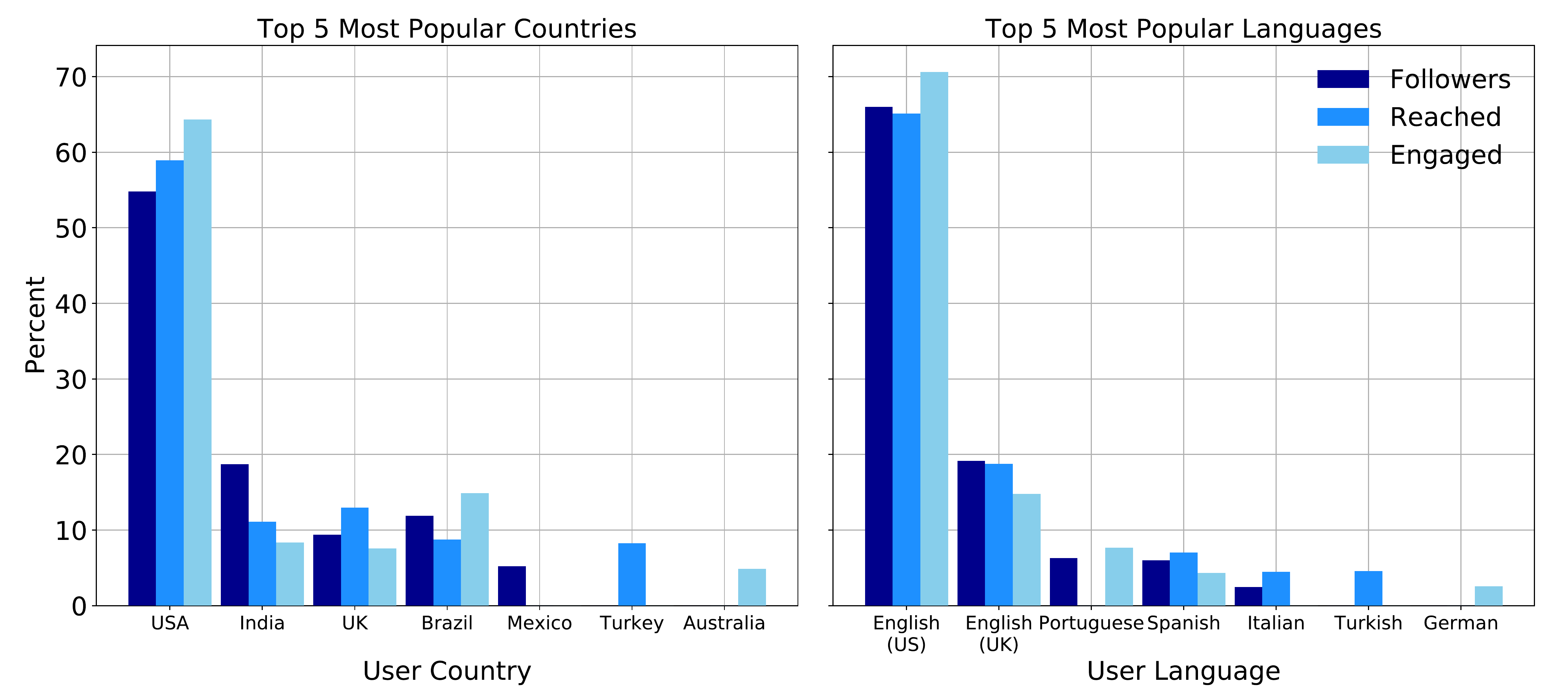}
\caption{Percent of Facebook followers (dark blue, $n=7914$), users reached (blue, $n \approx 15000 $) and engaged users (light blue, $n \approx 500$) in the five most popular user-identified countries and languages of origin.}
\label{fig:fbuser}
\end{figure}

Figures~\ref{fig:menwom} and~\ref{fig:fbuser} present Facebook demographic information for DES' three primary social media user groups, based on level of engagement. Users who have ``liked'' or ``followed'' the Facebook page are categorized as ``Followers,'' users who are exposed to DES social media posts (e.g., who simply see the post on their mobiles phones) as ``Reached,'' and users who actively engage with content (e.g., who comment on, or share, material) as ``Engaged."

In Figure~\ref{fig:menwom} we analyze each user group by age and gender (as gathered by Facebook). The DES Facebook follower base is roughly 75\% men and 25\% women $(n=7914)$. Of these, the {\it Engaged} users are roughly 60\% men and 40\% women $(n \approx 500)$. The largest user age group ($\approx 20\%$) is aged 25-34, for both men and women and across each of the three user groups. According to Twitter metrics, most of the Twitter followers have self-identified an interest in science. Ninety-four percent of DES followers express an interest in ``Science News'' and 89\% of DES followers express an interest in ``Space and Astronomy'' $(n = 1402)$. 

Figure~\ref{fig:fbuser} displays the five most popular user-identified countries of origin and languages of the Facebook users in the three user groups. As shown in Figure~\ref{fig:fbuser}, the majority of the Facebook users in each of the three primary user groups are located in the United States and speak English (US). Each user group also includes users from India, the UK, and Brazil. In addition to English (US), English (UK) and Spanish are both in the most popular languages for each user group. We note that the fifth most popular country for each user group is unique; the ``Followers'' are found in Mexico, the ``Reached'' in Turkey, and the ``Engaged'' in Australia. 

We find DES social media user demographic information, for the most part, unsurprising. While the majority of Facebook users in general are young women \cite{duggan2013}, the fact that the majority of DES followers are young men is consistent with known issues concerning women in astronomy and astrophysics \cite{NAP12062,Ceci,Ivie2005}: 1) lower documented interest, 2) lower academic retention rates, and 3) professional underrepresentation. Furthermore, DES social media primarily reaches people predisposed to an interest in science, and does not necessarily inspire new interest. According to these statistics, we have not yet succeeded in reaching and maintaining audiences underrepresented in STEM (Section~\ref{subsubec:DnI}).

As expected, most DES followers are in the United States and English speakers. Perhaps the most puzzling demographics are those describing where DES followers are located. We are unsure why one of primary locations of ``Followers,'' Mexico, is not included in the primary locations of people ``Reached'' or ``Engaged.'' We are also unsure why DES posts are particularly popular in Turkey. Investigating these demographics in further detail will be explored in a later publication. 

As shown in Figure~\ref{fig:fbuser}, Spanish-language speakers are in the top five the DES Facebook ``Followers,'' ``Reached,'' and ``Engaged'' user groups. Combined with the fact that DECam is based in a predominantly Spanish-speaking country, this evidence suggests that translating DES content to Spanish is a worthwhile use of EPO resources and supports GP7 (see Section~\ref{subsec:transla} for more discussion of the translation of EPO content from English). 



Based on these metrics, we find that although the number of people following DES on social media has increased, it has not increased at the rate we expected. We also conclude that although we may have increased awareness of the DES project, it appears the primary audience are those who already self-identify as having an interest in astronomy or science in general. It is unclear if these users are scientists themselves or enthusiastic members of the public. It is unlikely that we have inspired \textit{new} interest in astronomy. If other large collaborations/organizations planning EPO science programs seek to use social media to captivate an audience that is not already interested in science, we suggest 1) that the social media strategy be structured for a specific target audience and 2) there is an emphasis on social media products that require only a few seconds to read and understand (e.g., ``memes''), preferably with links to more in-depth materials.



\subsubsection{Other Online Platforms}


The discussion above is focused on Facebook and Twitter. However, DES EPO has also used other platforms, e.g., Tumblr\footnote{\tt www.tumblr.com} for Dark Energy Detectives (Section~\ref{subsubsec:DED}) and Flickr\footnote{\tt www.flickr.com} for image curation (Section~\ref{subsec:images}). We also rely on our collaboration website (Section~\ref{subsec:website}) as a mechanism to archive social media content, and to provide a repository of in-depth information that can be linked to social media.


\section{EPO Programming for a Collaboration}
\label{sec:programs}


 


\begin{figure}[htp]
\centering
\includegraphics[scale=0.6]{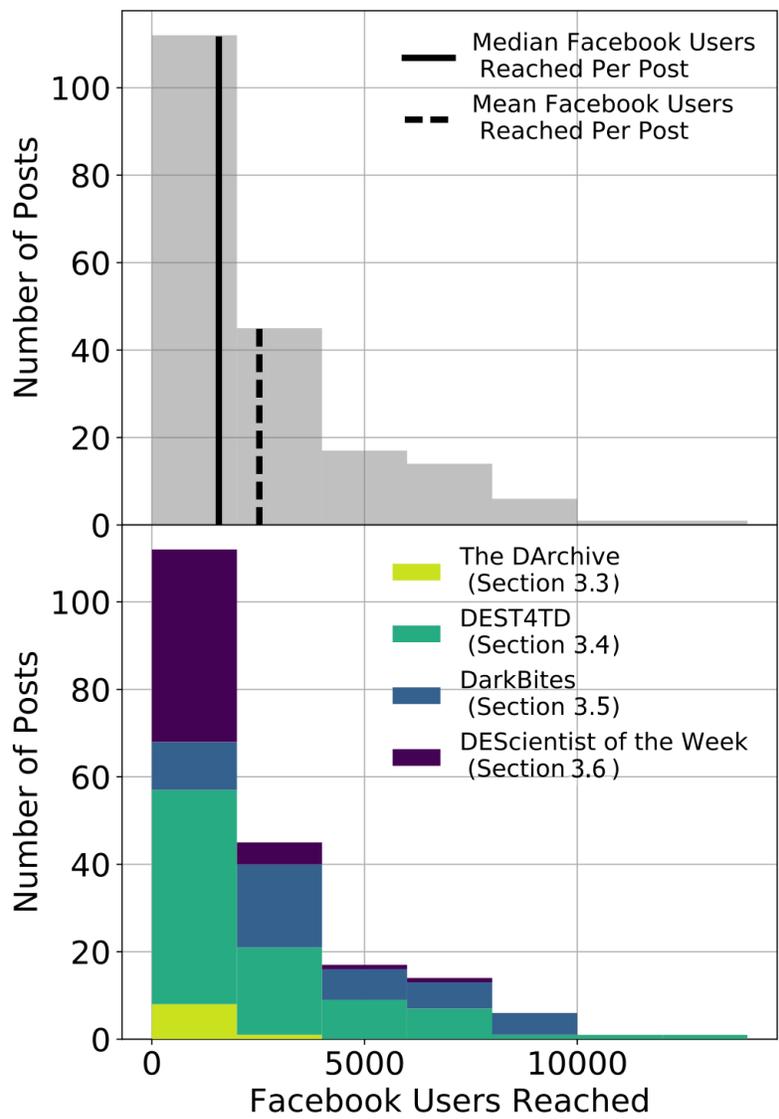}
\caption{Global distribution of Facebook likes per post for DES EPO products featured on social media described in this work (DArchive -- Section~\ref{subsec:darchive}; DEST4TD -- Section~\ref{subsec:t4td}; DarkBites -- Section~\ref{subsec:darkbites}; DEScientist of the Week~\ref{subsec:sow}). The mean reach (upper panel, dashed line) was 2,525 users; the median reach (upper panel, solid line) was 1,575 users. The bottom panel features a stacked histogram showing the number of posts per EPO product discussed in this section.}
\label{fig:totallikes}
\end{figure}

\begin{figure}[htp]
\centering
\includegraphics[scale=0.5]{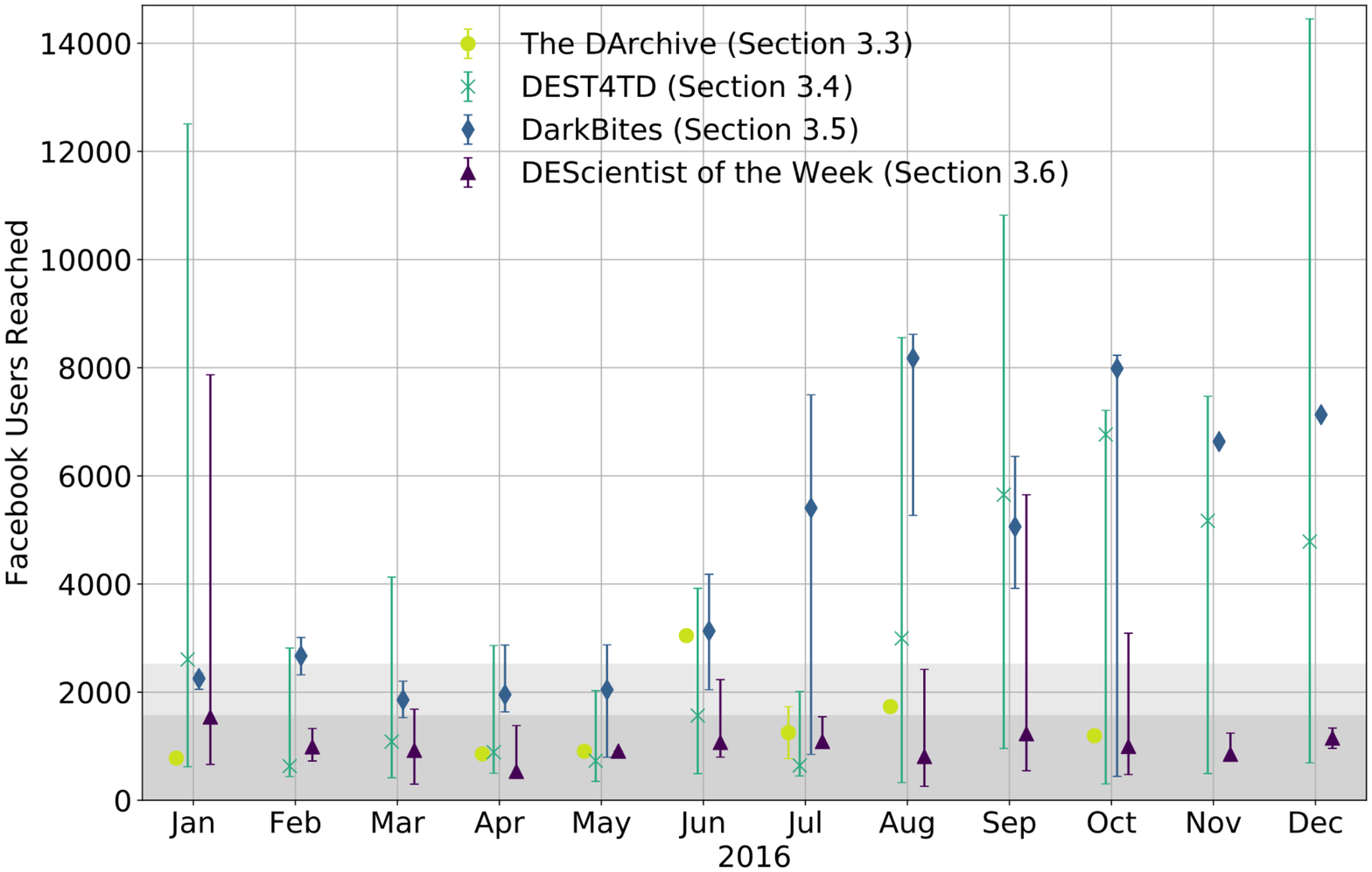}
\caption{Comparison of Facebook user reach for each of the social media DES EPO products featured in this section. Points represent the median reach of a project in a given month. Error bars represent the range of post reaches in a given month; i.e., no error bars indicate there was only a single post of that type. The height of the light gray bar signifies the average reach of all the posts from the 2016 calendar year (dashed line, upper panel of Figure~\ref{fig:totallikes}); the height of the dark gray bar signifies the median reach (solid line, upper panel of Figure~\ref{fig:totallikes}). We note that the general observed peak in activity around August-September corresponds to the beginning of the DES observing season.}
\label{fig:smsum}
\end{figure}

In this section, we discuss the primary online\footnote{These ``primary online'' initiatives comprise only a fraction of the entire repertoire of DES EPO activities.} DES EPO initiatives organized by the EPOC\footnote{Education \& Public Outreach Committee, as defined in Section~\ref{sec:desover}.} since Fall 2014. For each initiative, we present a summary of the project and discuss its implementation as well as project strengths and challenges. We also offer recommendations for EPO efforts connected to other large collaborations/organizations planning EPO science programs (e.g., LSST). Corresponding logic models detailing the specific inputs (e.g., time, funding), outputs (e.g., online content, lesson plans), and outcomes (e.g., number of participants, public and scientists' reactions to projects) for many of the initiatives presented here can be found in Appendix~\ref{sec:lm}. In some cases where project products were distributed via social media, we present specific Facebook reach statistics over the 2016 calendar year (we find the Twitter statistics roughly mirror these trends). We use reach statistics only from 2016 as there was the greatest temporal overlap between projects. We compare these metrics to the global reach of all DES products featured on social media (see Figures~\ref{fig:totallikes} and~\ref{fig:smsum}) as a benchmark for impact. We note that many of the projects described below are on-going, so the discussion presented here is limited to our experience until April 2017.  

\subsection{Dark Energy Detectives}\label{subsubsec:DED}


One significant EPO effort was in place before the formation of the EPOC in Fall 2014: Dark Energy Detectives (DED). This was a blog series, organized by one of the future founders of the EPOC, that used creative writing to tell stories related to the science, technology, and processes underpinning DES.  

In each DED entry, the writer assumed the role of a detective or sleuth and described a mystery in which cosmology was the backdrop. The style was borrowed from film noir and emulated many works of cinema and fiction that have permeated pop culture over the last 40 years. Each post featured an image or video related to DES. This strategy was employed in an effort to reach audiences that had not previously engaged with science EPO. DED was published via Tumblr,\footnote{\texttt{http://darkenergydetectives.tumblr.com/}} and reached $\sim$$45,000$ Tumblr followers at peak popularity. It was supported by a dedicated Twitter account\footnote{\texttt{https://twitter.com/darkenergydetec}} that also strictly maintained a film noir tone.




DED posts were made over the course of three DES observing seasons (August to February), with a bi-monthly cadence. During the first of these observing seasons, all of the posts were written and managed by a single individual and benefited from professional editorial assistance (from the Fermilab Communications' Public Information Officer). Thereafter, an increasing fraction of the posts were written by guest authors (other DES scientists), with editorial assistance provided. Each piece required one to three weeks of writing and editing (albeit not full time), including correspondence between writer and editor. The prescribed narrative tone required close coordination between the editor and each writer. 

After the formation of the EPOC, attention and resources switched focus from DED to providing support to other DES EPO activities. As a result DED was phased out after the end of the third observing season.




\subsubsection{Discussion}

By metrics of user reach, and the creative use of new media, DED was a successful DES EPO product. The DED project gave eager scientists the opportunity for creative writing. However, maintaining the film noir style amongst a community of writers (many of whom were unfamiliar with the genre) was difficult. Only a handful of DED posts were made after the formation of EPOC diverted volunteer time elsewhere. The project would have been more efficient and sustainable had it relied on just one or two dedicated writers with prior experience in science communication. 

\subsubsection{Take-Home Messages} 
Tumblr is an appropriate platform for long-form, blog-style EPO content (see Section~\ref{subsec:darchive} for a contrasting example). It is best to target scientists who enjoy writing for the public (and who have prior experience) for long-form pieces, rather than trying to encouraging everyone in a collaboration/organization to take part. That said, it would be a good use of a collaboration's/organization's financial resources to  provide training opportunities to all members; in doing so, scientists might discover a previously unrecognized enjoyment for, and/or talent in, science writing. 




\subsection{The {\tt darkenergysurvey.org} Website}
\label{subsec:website}

When the EPOC\footnote{Education \& Public Outreach Committee, as defined in Section~\ref{sec:desover}.} was founded in October 2014, NRW assumed the responsibility of updating the DES website. The aim was a modern, user-friendly, website with integrated social media (GP5, GP6; Section~\ref{subsec:hypo}).

Rather than update the content of the existing DES website, we decided to create a new website that would meet aesthetic, content, and user-interface goals. This involved updating the ``back-end" data access structure as well as the ``front-end" publicly accessible presentation layer. This process, from development to public launch, included: 1) seeking out an external web development agency and obtaining their advice, 2) designing the layout, user experience, and information content, 3) organizing the back-end features necessary for page updates and maintenance, 4) creating, reviewing, and formatting content, and 5) designing a strategy for website maintenance.

Key choices for the front-end revolved around the user experience. We believed that if this element was appealing, simple, and intuitive, the audience would be able to navigate easily to the online content, and want to return to it in the future. Our goal was to create a site that was easy to navigate, both on a computer and mobile device, for a variety of user groups (e.g. professional astronomers, educators, general public). After much internal discussion and research of existing science collaboration sites,\footnote{E.g. \texttt{hubblesite.org} and \texttt{sdss.org}} we opted for a compromise between a multi-page, hierarchical structure and one with more modern media-driven features. This allowed us to create a bridge between the past and present of science collaboration web pages; appealing to self-identified science enthusiasts who were already used to exploring well-organized and curated sites and new audience members who might be attracted by creative content and the multimedia-oriented main page. 

Implementation of these front-end features relied on an understanding  of back-end development. For reasons related to budget and site maintenance, we elected to utilize an existing template service\footnote{\texttt{https://wordpress.com/}} for the back-end of the new website. However, as the EPOC coordinators did not have the necessary web developing experience, we contracted a professional website development team to adjust the back-end structure to suit our needs. Funding for this project was approved by the DES director and was drawn from contributions from DES collaborating institutions (Appendix~\ref{subsec:org}). We worked with the development team for a year and a half and the cost ultimately came to $\approx\$5000$. This was the largest single-project budget of any of our other EPO programs and greater than the sum of EPO funding allocated for all other DES EPO projects.


Much of the developers' time was spent organizing the back-end structure so that site maintenance would be straightforward. For example, they created a slide interface for easy graphic uploading and multiple web forms for adding new content. Once the website was publicly launched, updating and adding content and other site maintenance was under the purview of the EPOC. Much of the content for the new site was transfered from the old; however, we devoted a substantial amount of time to updating and rewriting sections on a range of topics from DES science to collaboration structure. Most of the time from the EPOC went toward creating an interface that was navigable by both new and returning visitors, and relatively extensible, should we find a need for new sections of content. 

The second and third phases took place concurrently and took $\sim$6 months to complete (not because they were particularly big tasks, but because they had to fit in with our primary responsibilities, such as research, teaching etc.). Converting and developing new content slightly overlapped with the earlier design phases and required $\sim$$3$ months to complete.

\begin{figure}[htp]
\includegraphics[scale=0.32]{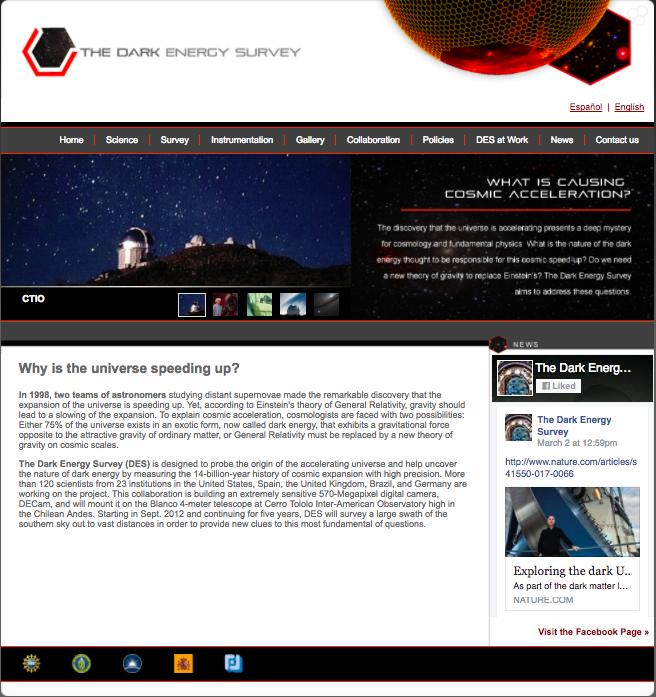}
\includegraphics[scale=0.31]{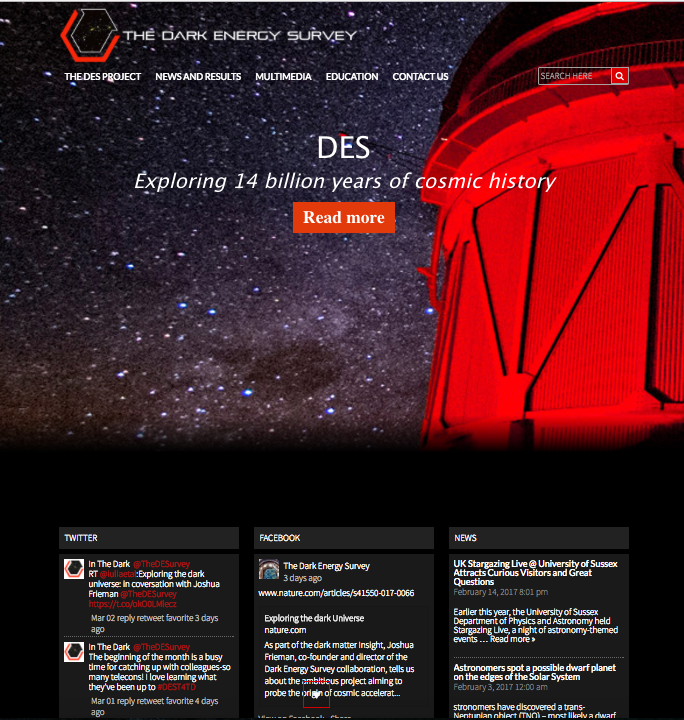}
\caption{Screenshots of original (left) and updated (right) DES website home pages. Note that the screenshot of the new page does not capture the full screen as the template is widescreen.}
\label{fig:homepage}
\end{figure}

Although we believe the current DES web page is a significant improvement from the previous public page (screenshots of the old and updated home pages are presented in Figure~\ref{fig:homepage}), development and maintenance required much more time and effort than we anticipated. In hindsight, we believe we devoted too much time to designing and creating the optimal aesthetic. This allocation of resources meant that we then did not have enough time to create and review static web page content or develop other EPO projects. In fact, several pages of written content had not been published (at the time of writing) because we lacked resources for editing. In addition, social media platforms have been taking up larger sections of the online landscape. These social media platforms house content and can provide gateways to an organization's website; in this paradigm, websites are most useful when they provide interactions and content that are not readily available through a social media platform. We believe that is the scenario in which content-driven science websites are most likely to be visited.

\subsubsection{Take-Home Messages}


A well-designed and well-maintained website is an invaluable resource for both the public and for members of the respective collaboration/organization. We strongly recommend other large collaborations/organizations planning EPO science programs devote whatever resources possible to contracting an external web development team who can lead back-end development and aesthetic design. By doing so, the time that volunteers can contribute can be used for writing scientific content. In our experience, the best way to generate content collectively is via ``hack sessions'' at collaboration meetings (emails requesting content from individuals did not usually deliver results). Finally we stress the importance of providing versions of the website in other (than English) languages (see Section~\ref{subsec:transla}).



\subsection{The DArchive: DES Results in a Nutshell}
\label{subsec:darchive}

The fundamental charge of DES is to conduct innovative, high-caliber research. As a large-scale ($>400$ members) science collaboration, DES scientists work together to produce new science results that are published in peer-reviewed academic journals. While members of the academic community know how to access and interpret these papers, they are not easily accessible to or digestible by the public due to the use of technical, scientific language. The ``DArchive: DES Results in a Nutshell'' project was designed with the goal of dissolving these barriers and making DES science more accessible to the public (GP3; Section~\ref{subsec:hypo}). The project name was inspired by the arXiv\footnote{\texttt{https://arxiv.org/}} (pronounced ``archive''), an online open-access repository where many physicists, astronomers, and astrophysicists post their scientific publications. Our intent was such that each DArchive would feature a summary of a DES paper, using language and analogies intended to connect a public audience with the science. 

The initial goal of the project was to have a complementary DArchive featured with the release of each DES paper. In an internal collaboration-wide survey,\footnote{See Appendix~\ref{sec:intersur} for the internal survey used for formative evaluation of the DES EPO program after its first year.} 91\% of respondents $(n = 69)$ indicated their support for the project as they believed it to be a worthwhile DES EPO effort. Between May 2015 and January 2017, there were 89 DES papers submitted to academic journals; yet there were only 15 published complementary DArchive articles\tnote . Published DArchive posts were featured both on the DES website and on the DES social media platforms.

\subsubsection{Project Organization and Implementation}

\begin{figure}[htp]
\centering
\includegraphics[scale=0.55]{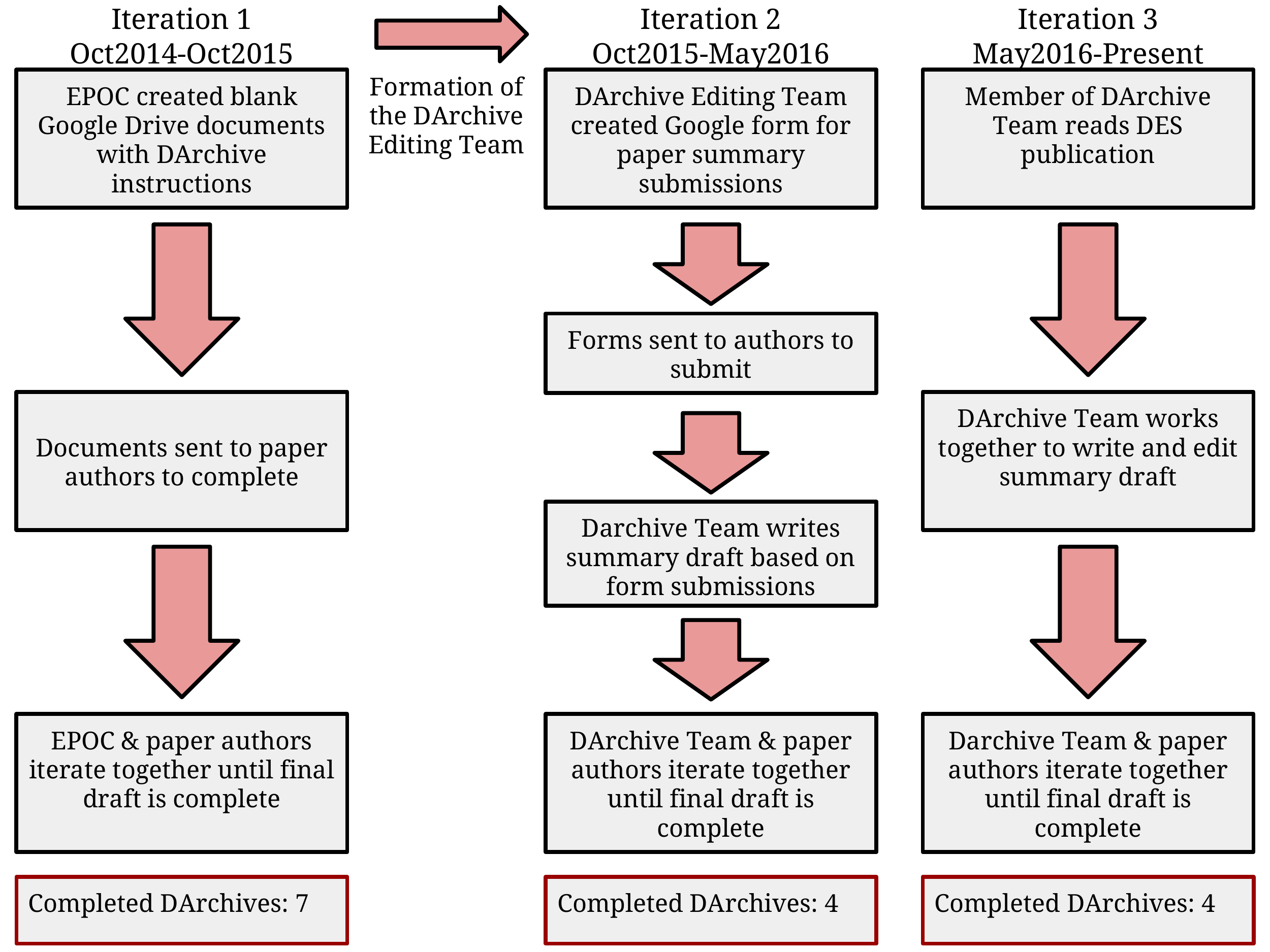}
\caption{DArchive work flow for each of the three project organization iterations.}
\label{fig:darchiveflow}
\end{figure}

Since its inception, the DArchive project has gone through three iterations of organization, which are summarized in Figure~\ref{fig:darchiveflow}. A logic model describing the inputs for each iteration is presented in Figure~\ref{fig:darchivelm}.
 
In Iteration 1, we expected the paper summaries would be primarily led by the paper authors (GP8). The EPOC\footnote{Education \& Public Outreach Committee, as defined in Section~\ref{sec:desover}.} created a DArchive template document, intended as a guide to help the authors draft their pieces, and sent this template, along with a request for a DArchive, to paper authors. Authors who responded to the request (typically graduate students or postdoctoral researchers) drafted a paper summary, and the EPOC worked with the authors to finalize the DArchive. 

During this iteration of the project, we encountered several key issues with author participation. Despite our efforts to get in contact with paper authors, the response rate, and willingness to participate was very low. We expected this might be due to the point at which during the publication process we contacted the authors, i.e., once the paper had been submitted for internal collaboration-wide review or after it had been published in a journal, but found no correlation with request timing and response rate. Some authors found it particularly difficult to translate their work for a public audience, and therefore the EPOC often had to spend hours editing a particular piece to meet the DArchive communication goals. Additionally, some paper authors did not understand the importance of eliminating or rephrasing some scientific terminology, and would argue with the EPOC about how to best convey a topic, significantly lengthening the DArchive writing and editing process. 

The goal of Iteration 2 was to decrease the amount of time and effort required by the paper authors. As part of this iteration, we assembled the DArchive Team: a group of five authors (including NRW) and one editor-in-chief. The goal of this iteration was to distribute the DArchive authorship and improve consistency and quality amongst posts. As part of Iteration 2, the DArchive team created a new DArchive submission form to help paper authors condense the most significant sections of their analyses. Questions on this form included: 
\begin{itemize}
\item{In one or two sentences describe the main hook of the paper.}
\item{In three or four sentences, describe your conclusions, results, and the reasons why you are excited about this work.}
\item{In a paragraph, describe how you came to these conclusions. Outline the main steps that lead to your results. Try to avoid too many technical details about systematic checks, etc.}
\item{In a paragraph, describe how your conclusions contribute to the advancement of knowledge about dark energy, cosmology etc.}
\end{itemize}
This form was sent to paper authors, and the DArchive Team drafted paper summaries based on authors' submissions. Unfortunately, the issues of Iteration 1 were also present in Iteration 2. The most significant problem was lack of participation from the paper authors. However, having a dedicated team of DArchive writers helped streamline the DArchive process and improved consistency between pieces. 

To further reduce the need for author participation, the DArchive Team adopted Iteration 3. Summaries in this iteration were driven primarily by the DArchive Team. A DArchive Team member would read a DES paper, write the paper summary, and then offer the DES paper author(s) a chance to include revisions. While the structure of this iteration gave the DArchive Team more autonomy, DArchive writers had a difficult time balancing the time commitment necessary to complete a DArchive summary with the demands of their other duties (research, teaching, administration, etc.). We found the total amount of time needed to complete a DArchive, from both the paper author(s) and the editing team, was roughly ten hours. This translated to the release of about one Darchive feature per month.

\subsubsection{Social Media Reach: The DArchive}
Figure~\ref{fig:smsum} presents the Facebook reach of DArchive-related posts in 2016. Note that while most of these statistics are derived from individual DArchive posts, general DArchive announcements, e.g., a link to the DArchive page on the DES website, also contribute. As shown in Figure~\ref{fig:smsum}, the number of people reached per DArchive-related post is variable, and is typically below the median number of users reached across all DES EPO social media products. The mean number of users reached is 1,310 while the median is 907. Figure~\ref{fig:smsum} also indicates that unlike the other DES EPO social media products, DArchive posts were featured only roughly once per month during seven months of the year.


\subsubsection{Discussion} 

As made evident during each iteration of the DArchive structure, the time commitment was the biggest obstacle blocking the project's success. It was challenging to get paper authors, who had already written an academic paper, and DArchive writers, who enjoy written science communication, to devote the time necessary for each DArchive summary. Throughout the iterations, the EPOC tried various approaches to incentivize participation in the project, including:
\begin{itemize}
\item{Asking the DES director to publicly support the project at collaboration meetings.}
\item{Requesting that the DArchive process be streamlined into the official publication policy.}
\item{Offering infrastructure credit towards data rights and DES Builder status for participation (see Appendix~\ref{subsec:org}).}
\end{itemize}

None of these approaches proved entirely successful. While the director offered public support for the project, the lack of official status within the publication policy made it difficult to encourage or motivate participation. We found the amount of time necessary to write and edit a polished piece was simply not realistic for full-time scientists who can only contribute to EPO on a voluntary basis. 

\subsubsection{Take-Home Messages}

Although an efficient DArchive strategy has not yet been reached and we have not published DArchives at our goal rate, experiences in each iteration were incredibly valuable. Overall, we learned that aiming for a DArchive summary for each DES paper was too ambitious, and likely unnecessary as occasionally papers would overlap with similar material (i.e., papers from ongoing analyses would build upon one another). We also learned that the background information needed to provide context for the scientific analyses was repetitive from piece to piece. We began writing static background articles (i.e., about concepts like redshift or gravitational lensing) which were shared on the DES website and intended to provide relevant links in each DArchive, but have not yet published them, due to lack of time. We found that having a ``DArchive Editor-in-Chief'' was essential, as it not only made the posts more consistent, but made the process easier for the writers and the paper authors.


In addition to providing an EPO output for the public, we also found the DArchive project had unexpected value for DES scientists. Writing these pieces gave their authors opportunity to improve upon their science communication skills. Moreover, DArchives make DES science results more accessible to DES members who were not directly involved. Rather than having to read the whole of an academic paper, DES members can read a DArchive piece and learn the salient background and results. Furthermore, should those DES members be invited to give a general DES presentation to peers or to the public, the DArchives can be used to effectively convey the information in a given paper to the audience.


We believe that projects like the DArchive involving high-quality science writing would greatly benefit from professional experience and dedicated funding. Rather than being written by full-time scientists, we recommend these pieces be written by a professional science writer. These pieces are important ``legacy'' content for DES, and are part of the static content on the DES website. We would also recommend finding a more impactful means of product distribution other than social media. The reach statistics presented in Figure~\ref{fig:smsum} are lower than other DES EPO projects. We conclude that long-form EPO content will have a greater reach if shared on Tumblr, rather than just on Facebook or Twitter (see Section~\ref{subsubsec:DED}). That said, this type of content should still archived on a collaboration's/organisation's website.

\subsection{DES Thought for the Day (DEST4TD)}
\label{subsec:t4td}

As described in Section~\ref{subsec:sm}, one of the first priorities of the EPOC\footnote{Education \& Public Outreach Committee, as defined in Section~\ref{sec:desover}.} was to revitalize the DES social media presence and create a vehicle by which we could increase public awareness of DES (GP5; Section~\ref{subsec:hypo}). NRW hypothesized that providing a regular stream of social media content would be the best way to engage an audience that might otherwise be unaware of or uninterested in DES. However, as working researchers, NRW did not have the time to be the sole generators of daily original content. Therefore, NRW asked collaboration members to contribute by submitting a DES Thought for the Day (DEST4TD): a short statement about their work, science interests, or daily routine. In addition to providing the EPOC with social media content, i.e., to post on Facebook and Twitter, DEST4TD also provided the opportunity for collaboration members to engage with the public using a new, unconventional medium which would not require much time or preparation (GP8).  

DEST4TD was also intended to serve as a channel through which we could share real-life experiences of DES scientists with the public and contribute to our long-term goal of making science and scientists more accessible to those outside academia (GP3) and those not typically represented in STEM fields (Section~\ref{subsubec:DnI}). 

We anticipated that the project would be well-received by the collaboration as it provided a means of reaching a large audience without requiring any previous experience. We also anticipated that collaboration members would respond to email requests for participation as they would be required to spend no more than five minutes on an individual DEST4TD. When asked about the project, 70\% ($n=30$) of collaboration members responded that they believed DEST4TD was a worthwhile DES EPO project.\footnote{See Appendix~\ref{sec:intersur} for the internal survey used for formative evaluation of the DES EPO program after its first year.} By April 2017, 127 different collaboration members contributed a DEST4TD; 13 members contributed more than five unique submissions. By comparison, only 3 DES members had contributed to DES social media prior to the creation of the EPOC.

\subsubsection{Project Organization and Implementation}


Developing the most effective project strategy required substantial trial and error in the early stages. Ultimately, we converged on a process whereby collaboration members were emailed each weekday using programming scripts scheduled on a server. Members receiving the emails were selected from the full collaboration member list (derived from a membership database) at random without replacement. Once the pool of remaining names had dropped below twenty, the full list was re-implemented. The subject line contained the member's first name to make the emails seem personalized. Those receiving the emails were asked to respond to a prompt that was selected at random from the following list:
\begin{itemize}
\item{This week for DES, I'm working on ...}
\item{The most exciting thing about working on DES is ...}
\item{The most difficult / frustrating thing about working on DES is ...}
\item{My favorite thing I've learned by working on this project is ...}
\item{The biggest mystery in [your specialism in DES] is ...}
\item{When I went observing for DES, I was surprised by ...}
\item{When I went to [my first, the most recent etc] DES collaboration meeting, I was surprised by ...}
\item{Submit a photo from observing or a figure from your research (if the latter, please make sure it has been approved for public dissemination)}
\end{itemize}
A longer list of suggestions was also included, in case the selected prompt did not inspire a response.




The response rate to DEST4TD email requests varied throughout the calendar year. Generally we received bursts of participation around the time of collaboration meetings and during the observing season (while scientists were at the telescope). As the project developed, the rate of response dropped off. At the time of writing, the strategy was to send ten requests per day to ensure an average response rate of one per day. Occasionally (roughly once per month) we receive spontaneous DEST4TD content, i.e., a submission not prompted by an email.

DEST4TD submissions were posted daily to the DES Facebook and Twitter accounts. This required: vetting of the source material, i.e., ensuring that any new science results were allowed to be publicly released and that submissions were sensitive to the diversity of cultures among the DES membership and DES social media followers; condensing posts to 140 characters for Twitter posting; finding related images and relevant article links. This process took $\sim$$5-10$ minutes per day. Although we could have added further automation by utilizing a social media dashboard, we opted to post manually as it was more convenient to devote time according to personal work schedules. Maintaining a regular posting schedule required a moderator who was confident managing several social media platforms and passionate about using social media as a vehicle for EPO.

A full description of the final DEST4TD framework is presented in Figure~\ref{fig:dest4tdlm}. 

\subsubsection{Social Media Reach: DEST4TD}

Reach statistics for DEST4TD are presented in Figure~\ref{fig:smsum}. Measured over a calendar year, the average reach per post was 2,649 users and the median was 1,567 users. We note that the spike in August occurs in the same month as the beginning of the DES observing season. As shown in Figure~\ref{fig:smsum}, the reach of DEST4TD posts is the most variable amongst the four DES EPO products. The minimum number of users reached was 303, while the maximum number of users reached was 14,450.  

\begin{figure}[htp]
\centering
\includegraphics[scale=0.47]{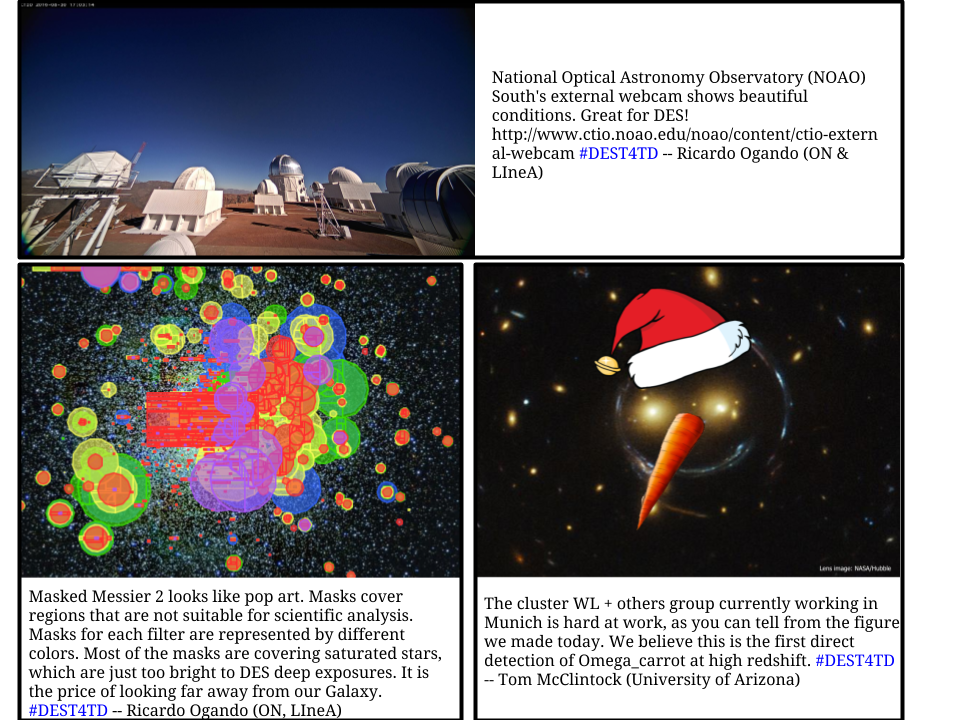}
\caption{Three most popular DEST4TD posts of 2016 submitted by collaboration members as determined by social media reach. Posts were submitted in January 2016 (bottom left), September 2016 (top), and December 2016 (bottom right), respectively.}
\label{fig:t4td}
\end{figure}

The three most popular DEST4TD submissions, as determined by reach on social media, are presented in Figure~\ref{fig:t4td}. As displayed in Figure~\ref{fig:t4td}, there does not seem to be a unifying theme amongst these most popular posts. The most popular post of January featured a masked image of a Messier galaxy that was presented in the context of Pop art. The most popular post of September featured an image of the Blanco telescope and surrounding instruments. One of the last posts of the year, and the most popular of 2016, featured a photo-shopped HST lens image, and a whimsical play on the Christmas holiday and cosmological parameter inference.

\subsubsection{Discussion}

The DEST4D initiative tapped into the creativity of hundreds of active scientists.  As a result, the posts have been characterized by freshness, authenticity, and (often) quirky humor. Since participating in DEST4TD required no previous experience, the project proved to be an effective tool for recruiting participation in DES EPO. 

Throughout the course of the project, we found that the popularity of a particular DEST4TD was rather unpredictable. Photographs colleagues submitted during or after observing, e.g., of the telescope site or flora or fauna on the mountain, were generally the most popular and reached thousands of social media users. We have thus far been unable to assess if DEST4TD has made any progress in better connecting DES science and scientists to the public. 

Perhaps some of the most valuable information we learned from DEST4TD was not about developing and distributing social media content, but rather about how scientists perceive themselves and their ability to engage in outreach on social media. In a DEST4TD follow-up survey, we asked collaboration members about DEST4TD emails and responses to learn how to improve project participation. While only 6\% responded that they reply to email requests right away, 20\% responded that they submit a DEST4TD once they felt a bit more inspired, and 30\% responded that they simply forget to respond to the request $(n=59)$. When asked why they did not respond to DEST4TD requests, 12\% of collaboration members noted that they ``did not have anything interesting to share.'' 


\subsubsection{Take-Home Messages}


We advocate that ``crowd sourced'' social-media based initiatives such as DEST4TD are a net positive when it comes to increasing scientists' public engagement.  They provide a steady stream to engage current social media ``Followers'' (Section~\ref{subsec:sm}) between official press releases describing major new science results. In addition, they give scientists who may otherwise have little or no EPO experience an opportunity to engage with a public audience without the need to overcome a steep learning curve. 

However, our experience developing and implementing DEST4TD illuminated some key issues that we had not anticipated at the outset: 1) social media is an excellent tool by which scientists can engage with the public, but reaching a target demographic is nontrivial; 2) scientists need to be reminded that the public are fascinated by the process of doing science, not just the final results; 3) providing daily original content is not sufficient to ``exponentially'' build a social media following (at the time of writing, the DES social media following was increasing by one or two users per week); 4) resources must be explicity allocated for engaging users who interact with (e.g., comment on) posted content; and 5) resources must be allocated to manage the infrastructure and to curate the best material in DES4TD posts (as these could have provided legacy content for DES).

With regard to issue 1), we suggest that specific social media strategies be developed to ensure posts reach target user groups. These strategies could be informed and evaluated by focus groups and market research. Sufficient resources (personnel and budgetary) for this process should be allocated from the outset. With regard to issue 5), the DEST4TD project heavily relied on automated emails and post vetting and management. One challenge was that these emails were frequently being returned to sender, often because the DES membership database was not kept up-to-date; at least 20\% of sent emails were returned. Another significant challenge was maintaining the posting content and schedule without the primary DES social media manager (e.g., if the manager were sick or on vacation). We recommend that future groups interested in pursuing a similar project invest resources into membership database management and find several people who can assume responsibility for automated emails and are confident in their social media abilities.


\subsection{DarkBites}
\label{subsec:darkbites}


The DarkBites project was inspired by the popularity of short, astronomy-related media that include analogies or ties to popular culture (GP3; Section~\ref{subsec:hypo}). In particular, we sought to emulate the style of well-known astrophysicist Neil deGrasse Tyson\footnote{\texttt{https://twitter.com/neiltyson}} and, as a result, reach a different audience to those following DEST4TD (Section~\ref{subsec:t4td}) or the DArchive (Section~\ref{subsec:darchive}). The initial concept was to generate short (one or two sentences), astronomy and cosmology sound bites that would surprise and inspire the public. The EPOC\footnote{Education \& Public Outreach Committee, as defined in Section~\ref{sec:desover}.} envisioned that these would be written in such a way that they would be accessible to anyone over the age of 10. At the Fall 2015 collaboration meeting, the EPOC identified a DES colleague (a graduate student) who was excited by the concept and was never short of ideas; he agreed to supply the text for the project. At about the same time, it came to NRW's attention that one of the DES postdoctoral researchers was a talented and prolific artist. NRW approached her to ask if she would be interested in providing occasional illustrations for the DarkBites project. She was genuinely delighted to do so, and created accompanying artwork for the weekly DarkBites text. Additional fact-generators and artists joined the project in the latter stages; the final DarkBites product consisted of 52 pieces of text and artwork.


\subsubsection{Project Organization and Implementation}

\begin{figure}[htp]
\centering
\includegraphics[scale=0.3]{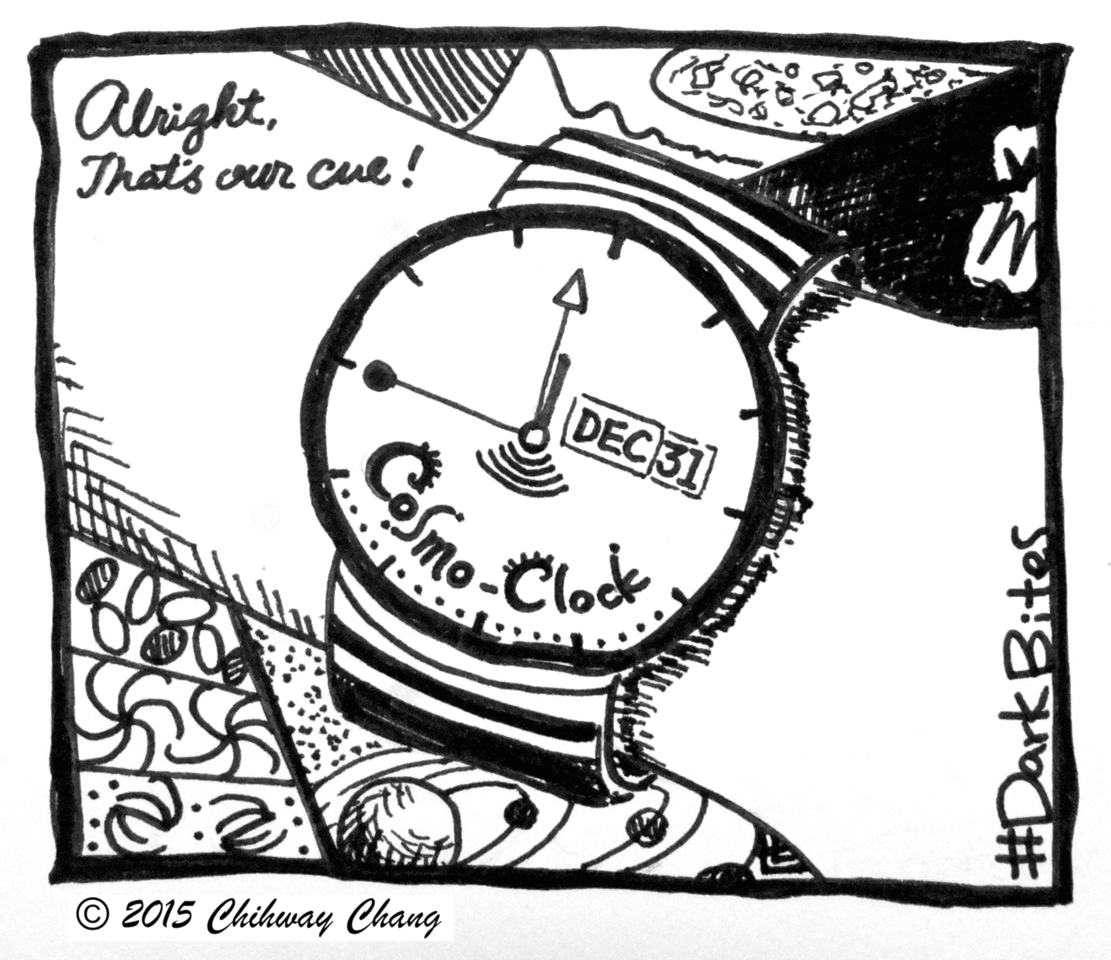} 
\caption{An example DarkBite image. The associated caption: If the lifetime of the universe were compressed to a single calendar year, the entirety of human history would occur in the last 15 seconds of December 31. Image Credit: Chihway Chang, University of Chicago; Fact Credit: Daniel Nagasawa, Texas A\&M University.}
\label{fig:db1}
\end{figure}

Participation in the DarkBites, which lasted roughly one year, was openly advertised to the entire collaboration.
Throughout the duration of the project, the DarkBites team was comprised of two dedicated fact-creators and three dedicated illustrators. Artistic decisions were made entirely based on the preferences of the project illustrators. There was no need for editorial input, which was fortunate as resources for input were insufficient. DarkBites facts were composed on a shared online document and illustrations were kept in a shared online folder. Over the course of a year, DarkBites were posted on social media and the DES website. DarkBites were archived on their own section of the DES website, and therefore could easily be used for subsequent posts and projects. Figure~\ref{fig:db1} features an example DarkBite image and caption\tnote . A more complete description of the project inputs and outputs is presented in Figure~\ref{fig:darkbiteslm}.


\subsubsection{Social Media Reach: DarkBites}

The number of users reached on Facebook by DarkBites posts is featured in Figure~\ref{fig:smsum}. As shown in the Figure, the median number of users reached was fairly consistent from January-June, and roughly plateaued from August-December. Over the course of the year, the mean number of users reached was 3,843 and the median was 2,873. If we consider only January-June, the mean and median number of users reached were 2,320 and 2,202, respectively; for the latter half of the year, the mean and median number of users reached were 5,800 and 6,360, respectively. 


We believe the increase in DarkBites reach after May could be attributed to two factors. At that time, two new illustrators joined the project and we began featuring images in color. It is also possible that this increase in reach was just a natural buildup of project followers over time. As shown in Figure~\ref{fig:smsum}, the highest-reaching DarkBites of 2016 were featured during the month of August. These two DarkBites were both related to sports, with the most popular referencing the ongoing Rio 2016 Olympics. Figure~\ref{fig:smsum} clearly shows that DarkBites posts were consistently some of our most popular on social media. The average and median reach of DarkBites posts were roughly equivalent to or greater than the global average EPO product reach.

\subsubsection{Discussion}


Perhaps the most valuable experience from the DarkBites project was witnessing the effect and impact of collaborative creativity. This project would never have succeeded without the artistic talents, creativity, and enthusiasm of our colleagues. We also observed how combining science with popular culture can be an effective tool for science communication. The most popular DarkBites posts of 2016 featured references to topical world events, suggesting that integrating science with ``trending'' topics may increase content reach.

\subsubsection{Take-Home Messages}

We found that a powerful way to inspire others to engage in EPO is to appeal to their personal hobbies or interests. We highly recommend that future EPO collaborations/organizations consider projects that capitalize on the talents of collaboration members.

Furthermore, the DarkBites project became a branching point for other EPO projects. Finished DarkBites were used by members of the EPOC to complement formal education curricula for elementary school children and in outreach events to inspire children to draw their own DarkBites images. We also designed a follow-up project, the ``DarkBites Unplugged,'' as a vehicle to further explain and define the astronomical information included in the original DarkBites. 

Through the organic evolution of the project came the surprising added benefit of its versatility. Products such as DarkBites which can be used in a variety of venues have become a valuable asset for our EPO repertoire. They have allowed us to develop additional activities using the same content, thus leveraging the initial investment of the time involved to develop DarkBites. In hindsight, we would have designed more projects with this in mind and recommend other projects consider this in the future. 




\subsection{DEScientist of the Week}
\label{subsec:sow}

The DEScientist of the Week initiative was designed to increase scientists' accessibility to the general public (GP3, GP4; Section~\ref{subsec:hypo}). First and foremost, we sought to highlight the diversity in race, gender, and personality of collaboration members and scientists in general (Section~\ref{subsubec:DnI}). We also wanted to provide our colleagues with a means of speaking openly and honestly about their experience as professional researchers (GP8).

\subsubsection{Project Organization and Implementation}
Each DEScientist of the Week piece featured a profile of a randomly selected DES collaboration member. This profile included a photograph (if desired), a small summary of research interests, and a short-form interview. Interviews were conducted using an online survey that included questions such as:
\begin{itemize}
\item{What is your favorite part about being a scientist?}
\item{When did you know you wanted to be a scientist?}
\item{Do you have any hobbies or play any sports?}
\item{What motivates/inspires you?}
\item{If you weren't a scientist, what would your dream job be?}
\item{Any advice for aspiring scientists?}
\end{itemize}
The profiles were posted weekly on their own subpage of the DES website\tnote\,and linked on social media. A complete description of the project inputs and outputs is presented in Figure~\ref{fig:desowlm}.

\subsubsection{Social Media Reach: DEScientist of the Week}
Figure~\ref{fig:smsum} presents the Facebook reach of DEScientist of the Week posts. The average number of users reached was 1,285; the median number reached was 977. As shown in Figure~\ref{fig:smsum}, most DEScientist of the Week posts consistently reached less than the DES EPO global median and mean number of users. The post with the highest reach was featured in January and highlighted a female collaboration member from a DES institution in the United Kingdom. This post was not the first of the year, nor the first feature of a female scientist. This profile was however featured the day after one of our highest reaching DEST4TD posts (Section~\ref{subsec:t4td}). 



\subsubsection{Discussion}

Thus far, we have featured over 80 scientists as DEScientist of the Week. Initially, we asked only those who self-selected for the internal DES EPO LISTSERV (electronic mailing list) to participate and the response rate was nearly 100\%. After opening the project to the rest of the collaboration, the total response rate was 60\% $(n=149)$, despite the fact that 82\% $(n=85)$ of surveyed\footnote{See Appendix~\ref{sec:intersur} for the internal survey used for formative evaluation of the DES EPO program after its first year.} collaboration members responded that they believed project was a worthwhile EPO effort. This response rate is much higher than that for DEST4TD (Section~\ref{subsec:t4td}), which varied between 10\% and 20\% depending on the time of year. We believe this higher response rate was primarily due to the fact that while scientists may feel apprehensive about talking about their research with the public, they feel flattered when asked to talk about themselves. Some collaboration members have even asked to be featured as DEScientist of the Week to coincide with job applications. 

\subsubsection{Take-Home Messages}

As our colleagues generally have responded favorably to the project and it did not require a great amount of administrative effort, we recommend facilitating similar scientist profile/interview initiatives, as they highlight the diversity in the scientific community and they have potential for significant long term-impact. The infrastructure of large collaborations/organizations is well-suited for this project, and we believe extensions to specific university departments or any individual with access to a network of scientists would also be impactful. Given our observed response rate, we suggest polling a network of roughly 200 scientists to create a sizable repository. 

We continue to work toward the long-term goal of changing public opinion of science and scientists, which may ultimately include updating the structure of the project. As demonstrated in Figure~\ref{fig:smsum}, DEScientist of the Week posts did not reach as many followers as other DES EPO initiatives. This suggests that the current structure may not be the most effective strategy for content distribution. However, interactions with DES social media users, including shares, likes, and comments, lead us to believe we are indeed making an impact, even if only on the smallest scales. For example, social media users have commented that find the pieces inspiring. 


\subsection{Multilingual EPO}
\label{subsec:transla}

As DES is an international collaboration of scientists from seven different countries and a survey relying on data from the Blanco Telescope at CTIO,\footnote{\texttt{http://www.ctio.noao.edu/noao/}} we sought to allocate EPOC\footnote{Education \& Public Outreach Committee, as defined in Section~\ref{sec:desover}.} resources towards projects that we believed would reach the broader, international community (GP7; Section~\ref{subsec:hypo}). Without funding to pay for translation services, we relied on volunteers in the collaboration to meet this goal. We were fortunate that several bilingual collaboration members, particularly Spanish speakers, are passionate about communicating science in languages other than English. While there has been some original content created for international audiences, most of this effort focused on translating existing DES EPO content. 

\subsubsection{Project Organization and Implementation}

\paragraph*{Translation of website --} The most ambitious of these translation projects was a full translation of the DES website, including sub-page content, into Spanish. There is now a Spanish version of the DES website. To allow this version to be embedded within the English version, professional web-development expertise was needed, and this required an investment of $\approx$$\$1000$. Translations were done by a group of four native Spanish speakers in the collaboration who volunteered to translate website content.

\paragraph*{Many Languages of DES --} We attempted to geographically broaden the DES reach via the ``Many Languages of DES,'' or MLDES, initiative. In this project, we asked bilingual members of the collaboration to translate DarkBites (Section~\ref{subsec:darkbites}), DEST4TD (Section~\ref{subsec:t4td}), or other short-form, online DES EPO content on a weekly basis. Throughout the duration of the project, translated languages included: Spanish, Portuguese, French, Italian, German, Chinese, Farsi, Russian, Greek, and Serbian. Translators were reminded weekly to submit any translations by Friday afternoons. Translations were generally posted on social media on Sundays. The posting strategy for MLDES evolved throughout the lifetime of the project, and as a result, there was no straightforward mechanism to obtain reach statistics over time like those presented in Figures~\ref{fig:totallikes} and~\ref{fig:smsum}.
 
\paragraph*{Miscellaneous --} Several other DES multilingual initiatives emerged organically from volunteers. For example, some DES colleagues based in, or native to, non-English-speaking countries frequently produced DES-related content in their native languages, e.g., via writing press releases for their institutes (and then working with local journalists who see the releases) or posting about DES on their personal social media accounts (with content they produced themselves and/or translations of content produced by the EPOC).  Another small group of collaboration members manage an official DES account on Weibo, a Chinese hybrid of Twitter and Facebook. Weibo posts are roughly 40\% original content (loosely based on existing DES posts) and 60\% direct translations of DEST4TD or DarkBites. The Weibo page has roughly 2,000 followers and the average post reaches 1,000 users. The Weibo post with the largest reach ($\approx$$30,000$) featured an image of an Einstein ring discovered by DES reminiscent of the logo for Youku (Chinese YouTube). 

\subsubsection{Discussion}

\paragraph*{Translation of website --} This was a worthwhile endeavor. It was also quite straightforward to manage because bilingual volunteers were passionate about the project, volunteered many hours of their personal time, and required very little input from the EPOC. Indeed, our only regret is that we did not embed a Spanish version of the website from the outset.


\paragraph*{Many Languages of DES --} The translated posts did not have a large reach ($\approx$$200$ users per post). For this reason, the project was terminated in Fall 2016. We speculate that this was because the posts appeared on primarily English language distribution platforms, i.e., which had a user base that was already proficient in English, even if as a second language. In hindsight, we should have asked the MLDES translators to set up DES social media accounts in their respective languages.

\paragraph*{Miscellaneous --} The DES Weibo account was initiated spontaneously by a Chinese member of DES, and then enthusiastically supported by a handful of her compatriots. Unlike MLDES, this was a social media account designed specifically for a particular language market and consistently had a higher reach. Similarly, DES colleagues with a strong personal social media following in their native language have generally had a better response to their DES related posts than did MLDES.

\subsubsection{Take-Home Messages}


The wealth of translated content now available is entirely due to the participation and leadership of bilingual volunteers.
Translation initiatives are an excellent example of designing projects which utilize skills and talents of collaboration members. We recommend that other large collaborations/organizations planning EPO science programs plan for multilingual versions of their webpages from the outset. We further recommend that if a collaboration/organization wishes to foster an international DES audience using social media that they maintain separate accounts for the different target audiences. 


\subsection{Image \& Video Curation and Creation}
\label{subsec:images}

Images and figures are a vital component of DES science and DES EPO, and are used for science communication both within the scientific community and with the public. As part of the goal to centralize DES EPO efforts and products, we attempted to consolidate image management for the collaboration as a whole (GP9; Section~\ref{subsec:hypo}). This was more successful for some types of images than others. In this section we discuss the creation and curation of four main types of images: 1) figures that appear in DES publications and talks, 2) false-color DECam (astronomy) images, 3) photographs and simple videos, 4) astrophotography, infographics, and edited videos.


\subsubsection{Project Organization and Implementation}
\paragraph*{Figures that appear in DES publications and talks --}

Any publicly-released DES material (e.g., figures, refereed journal articles, or DES-approved conference presentations) can be used by DES members for other purposes, including: talks or presentations (for fellow astronomers/scientists or for the general public), popular science articles, or social media posts. To ease the process of accessing figures approved for public dissemination, the DES Publication Board created a central \emph{Figure Library}. This is a password protected-database of selected plots and accompanying descriptions; descriptions are written for DES members who are not experts in the relevant subfield. The collaboration was initially slow to adopt the \emph{Figure Library}. However, after its use became part of the official DES publication policy, compliance has increased significantly. By April 2017, the DES \emph{Figure Library} had become a useful resource for DES EPO.


\paragraph*{False-Color DECam Images --}
False-color\footnote{All astronomy images taken by CCDs are gray scale. However, if several images are made of the same patch of sky through different filters, then it is possible reconstruct a color image using software.} images of astronomical objects are vital for effective communication to the public. This is best demonstrated by the huge popularity of the images in the Hubble Space Telescope gallery. For many years, the DES data management (DESDM) team did not prioritize the development of software to generate these high quality-images (which are not often necessary for scientific analyses). Instead, DESDM primarily devoted their limited resources to creating the source catalogs necessary for DES science. To fill the gap, a handful of collaboration-member teams unilaterally developed their own software for false-color image generation. These teams were largely motivated by the needs of their scientific research, but some were instead (or also) motivated by EPO goals. All of these efforts were well-intentioned, but there was significant overlap and duplication. The resulting lack of cohesion, especially with regard to agreed quality standards, meant that the EPOC\footnote{Education \& Public Outreach Committee, as defined in Section~\ref{sec:desover}.} was not able to fully exploit the vast wealth of DES image data.



\paragraph*{Photographs and Simple Videos --}

The reach and impact of the online initiatives was significantly enhanced when pictures or other visual media were included. The DES EPO program was fortunate that it did not rely on choreographed  ``photo-ops'' or professional photographs. Rather, we capitalized on the fact that collaboration members like to take, and share, photographs and videos related to their science experience (e.g., from DES observing trips, DES collaboration meetings, or working group meetings). 
The DES EPOC regularly encouraged members to share their images, which were then archived using a private \emph{Flickr}\tnote\ repository. Meta-data, including photographer name, image location, and photograph date, were collected with the images. Given that an individual member may generate hundreds of photographs during a single observing shift, we selected a small sample of images to embed on the DES website. Videos embedded on the website were selected in a similar fashion; additional video material is shared via the DES Youtube channel\tnote. Collaboration members who contributed visual media were encouraged to sign a Creative Commons agreement, as this eased the process of distribution to outside media sources (e.g., local newspapers). The DES \emph{Flickr} and YouTube accounts were maintained by DES members on a voluntary basis.

\paragraph*{Astrophotography, Infographics and Edited Videos --}


Advances in technology have enabled professionals and amateurs alike to capture pictures and videos of the night sky. However, it requires technical skill, and expensive equipment, to produce astrophotography images and videos.\footnote{Astrophotography is a subdiscipline in amateur astronomy that seeks to produce aesthetically pleasing images rather than scientific data.} The production of time-lapse videos and images from these photographs is also technically difficult and time consuming. Astrophotography images and videos were of particular value to DES EPO, as they capture the public's imagination in a way that even false-color DECam images cannot.

Infographics\footnote{Graphic visual representations of information, data or knowledge intended to present information quickly and clearly.} are another genre of image that were integrated into DES EPO. Infographics can be static, moving (i.e., animated .gif format), and/or interactive. Compared to snapshot photographs, creating infographics requires more technical complexity, design considerations, and knowledge of the target audience. The DES EPOC commissioned (from volunteer members) a handful of infographics. For example, we enlisted volunteers to create an interactive world map showing the global distribution of DES members' host nations; this was useful not only for internal membership, but for media searching for scientists in their local communities. We also curated infographics developed by DES institutions for press releases or by individual DES members for their personal social media accounts.

Edited videos were the most difficult and costly (in terms of members' time) visual media to produce. The DES EPOC commissioned (from volunteer members) a handful of such videos. We also curated edited videos developed by DES institutions to accompany press releases.

Astrophotography, infographics and edited videos that were provided to the DES EPOC by DES members were archived and shared in the same way as simple photographs and videos (see above).



\subsubsection{Discussion}

\paragraph*{Figures that appear in DES publications and talks --} The DES {\it Figure Library} was developed after the formation of, and without consultation with, the DES EPOC. The lack of consultation, and the fact that the DES membership were originally resistant to the concept of a {\it Figure Library}, was for a while problematic for DES EPO. For example, the lack of compliance with the {\it Figure Library} meant that EPOC volunteers needed to extract figures from certain papers themselves to ensure the DES website publication pages were kept up-to-date.


\paragraph*{False-Color DECam Images --} Had the production of such images been prioritized by DES management, it could have been easy to have added a ``DES Picture of the Day''\footnote{Along the lines of the hugely popular Astronomy Picture of the Day ({\tt apod.nasa.gov}) initiative.} component to our online EPO portfolio.

\paragraph*{Photographs and Simple Videos --} In hindsight, we should have dedicate more resources (in terms of volunteer time) to the collection and curation of these types of data. There are thousands of member-produced images that have not yet to be loaded into the \textit{Flikr} repository. There were also several occasions when the media contacted DES asking to publish images and we were unable to work quickly enough to meet their requests, especially with regard to strict requirements on copyright. 

\paragraph*{Astrophotography, Infographics and Edited Videos --} The EPOC would have liked to have commissioned much more of this type of material, but this was not possible given the limited financial resources available.

\subsubsection{Take-Home Messages}

Media in the form of properly curated false-color DECam images, analysis figures, photographs, and videos are one of the most important legacy components of large scientific collaborations. However, in DES we have yet to fully exploit our potential to deliver interesting, eye-catching, and educational media. This is partly due to lack of resources, and partly due to lack of planning. We recommend that future large science projects design an image organization scheme early on and advertise image and video repository structure clearly and consistently. In hindsight, we should have put more time into the curation of DES members' photography. Much of that curation work could be done by undergraduate ``work-study'' students, and we encourage other collaborations/organizations to set aside a small financial sum to cover that type of work. We conclude that a centralized {\it Figure Library} to archive science figures from publications can be worthwhile, but that it needs to be included in a publication policy sequence from the outset to ensure compliance. Finally, and in the particular case of large-scale astronomy projects, we stress the need to generate eye-catching false-color images, and recommend that resources (including financial) are allocated, and quality standards agreed, centrally from the outset.

\section{Conclusions}
\label{sec:conclusion}

In this article, we described the strategy and programming of online Education and Public Outreach (EPO) in the Dark Energy Survey (DES). Unlike many other current large-scale optical astronomy EPO initiatives, DES EPO was not based on published data products, but rather inspired by the foundational science, data processing pipelines, and community of scientists that comprise the DES collaboration (Figure~\ref{fig:epochart}). Throughout this article, we detailed the strengths and weaknesses of a variety of DES EPO initiatives and provided recommendations, where appropriate, for other large collaborations/organizations planning EPO science programs.



Through the grass-roots approach of the DES EPO program, we generated a suite of novel and innovative online astronomy EPO initiatives. This was largely due to the passion and creativity of collaboration members eager to both contribute to Education \& Public Outreach Committee (EPOC)-organized activities and to develop new material on their own. The creation of legacy content, such as the DArchive summaries, in addition to more dynamic social media projects, such as DES Thought for the Day, provided a wealth of opportunity for scientists to connect with the public and with others in the scientific community. These activities would have perhaps been even more effective (in terms of content reach and scientist participation) had we focused our efforts on a few key projects rather than continue to develop and launch new initiatives. Upon reflection, the summative take-home message of the DES EPO experience thus far is to concentrate resources and effort on a select, smaller group of projects. However, it is important to recognize that it was only through the exploration of such a variety of initiatives that we were able conclude which projects which more deserving of limited effort and resources.

One of the most significant obstacles faced by the DES EPO program was the lack of time the EPOC coordinators and community members could devote to DES EPO programming. Given the amount of time necessary to complete projects such as a DArchive summary (Section~\ref{subsec:darchive}), the participation rate was lower than predicted from our polling of collaboration members' views. This mirrors other studies that have shown that generally scientists are in favor of EPO \citep{Ecklund,Andrews,Poliakoff}, but that their motives for and deterrents from participation in EPO remain unclear. For example, \citet{Dang} assert that barriers to astronomers' participation in EPO include 1) the perceived academic cultural norm that one can only spend time on EPO after completing necessary academic duties \citep{Ecklund} and 2) the lack of financial (institutional and grant) support for EPO. However, \citet{Dang} also find that astronomers in particular are widely supportive of EPO.  We will further explore DES members' EPO engagement practices and personal EPO motives and/or barriers in a future publication \citep{DESAttitudes}.



Many of the online DES EPO projects thus far have focused, at least in part, on changing the public perception of science and scientists. It is clear that this goal of changing the status quo is well-founded. Despite more recent integration of science into popular culture, e.g., books, television shows, movies, public perception of the ``typical scientist'' remains outdated. Fifth grade students largely still imagine scientists as ``white, eccentric males'' who work in a laboratory \citep{Barman}. This perception is generally similar for many adults, as the media tends to give very little coverage to scientists or the scientific method, making scientists seem removed from the general public \citep{Losh}. We hoped that by making scientists less ``distant" from the public that we might positively affect perceptions of scientists and their careers \citep{Losh}. We leave a long-term assessment of this aspect of the DES EPO program for analysis in a future publication.

Reflecting on our experience holistically, we believe that many of our initial guiding principles were not realistic given the collaboration organization and structure when the EPOC was created. Although the EPOC coordinators and many of the EPOC community members share an enthusiasm for science education and outreach, it is clear that GP1 and GP2 (Section~\ref{subsec:hypo}) have not been embedded into collaboration culture. Obviously we cannot expect that every collaboration member will share our goals, but perhaps officially integrating EPO into collaboration structure at the onset would have impacted general attitudes towards public engagement. Such structure may have also improved our experience trying to coordinate and incentivize EPO efforts (GP8, GP9). With regard to GP9, over the last three years we have noticed that some DES members with a passion for EPO continue to carry out their EPO activities unilaterally, despite repeated requests to  coordinate those activities with those being organized and implemented by the EPOC. Some of these unilateral initiatives have been of exceptional quality, including info-graphics (still and video) and articles for major media outlets. However, the lack of coordination has, on occasion, lead to a dilution of the DES EPO brand and negatively impacted the motivation of the EPOC.


We draw the following conclusions and make the following recommendations for others looking to pursue similar EPO endeavors:
\begin{enumerate}

\item{\emph{Integrate EPO into collaboration structure, policy, and culture.}
As described throughout this article, we encountered several issues that could have been avoided had DES EPO been more well integrated into the DES collaboration structure (see Appendix~\ref{sec:sciproj} for organization details). For example, the publication policy could outline that submission of each academic paper be accompanied by a short description and public-friendly figure which could then be featured on various EPO platforms. Although we received vocal support from the DES director, lack of official DES policy regarding EPO resulted in duplicated or inefficient EPO efforts and missed opportunities for targeted EPO projects for specific science results.}

\item{\emph{Establish an EPO budget before program development.} 
The EPOC was formed under the assumption that funding for DES EPO activities would be quite limited. However, an annual budget was never explicitly established. We requested funding as needs arose (e.g., for the development of the DES website). This system meant that we were never guaranteed funding for an EPO program. One area where we were unable to secure funding was to engage professional science writers to assist with the DArchive project. If we instead had firmly established a budget upfront, then we may have allocated resources differently and more strategically, i.e., ensuring undertaken projects reached their full potential (even at the expense of starting new projects).}

\item{\emph{Develop EPO initiatives with a range of required time commitments to increase the likelihood of scientist participation.} Collaboration members can only devote a small fraction of their time to organizing and engaging in EPO. However, designing, managing, and evaluating EPO programs requires a substantial amount of time. Unfortunately, spending time engaging in EPO remains undervalued in the astronomy community. This is reflected not only in personal attitudes but in funding. While we have created many innovative DES EPO programs, the list of projects that we have either not completed or not even started is much longer. This may be due to the fact that some of our goals were unrealistic and that we equated ``success'' with $\sim$$100\%$ scientist participation. We believed we could incentivize participation by offering credit towards data rights to collaboration members, but this did not prove sufficient; in the end most of substantive EPO work (i.e., not DEST4TD) was carried out by DES members who self-selected as EPO supporters. It was difficult to mobilize scientists to participate in EPO programs if they were not already inclined to do so.
}

\item{\emph{Contract professionals if budget allows. If there is not adequate funding, set limits on how much time can be dedicated to a particular project.} 
Contracting web developers for the DES website was essential. Not only did the EPOC not have the necessary experience to build a website, we also did not have the time. The time we did allocate to website development was often misplaced, as we focused on aspects such as structure and aesthetics rather than content development, where our science expertise was most relevant. When developing an EPO project, we recommend reflecting on how your expertise as a scientist will be utilized. If pieces of the project can be accomplished without your constant support, and if budget allows, consider external contractors for the task. If funding is unavailable, prioritize projects which balance time commitment and necessary experience.}

\item{\emph{Identify a specific target audience and methods for reaching that audience for each program.} Many of the discussed DES EPO projects were designed to inspire a general public interest in science. We hoped our online initiatives would reach people who might not otherwise be inclined towards the STEM disciplines. As social media metrics revealed, we were not successful in reaching this target audience. We encourage future EPO coordinators to think critically about content distribution and how to best interact with the desired audience. For example, we found that Tumblr (Section~\ref{subsubsec:DED}) had a larger reach for long-form content, and that Weibo (Section~\ref{subsec:transla}) was a more effective means to reach and international audience, than posting directly to English-based social media accounts. We also speculate that the DES EPO effort may have been more successful if we had explored other social media platforms, e.g., Instagram\footnote{\texttt{https://www.instagram.com/?hl=en}} and Snapchat.\footnote{\texttt{https://www.snapchat.com/}}}


\item{\emph{Inspire other scientists to participate in EPO by designing programs that utilize their personal interests and skills.} Some of the more innovative DES EPO projects, e.g., DarkBites, evolved from colleagues' artistry and creativity. We also found that collaboration members were more inclined to participate in projects which appealed to their personal interests.  This included projects centered around graphic design, astrophotography, video editing, and written science communication. We also believe these projects garnered more participation because colleagues could dedicate as much or as little time as they pleased, without feeling pressure to complete a project within a deadline. If EPO activities can align with scientists' interests, then perhaps time which might otherwise be spent on hobbies could be shared with EPO.
}

\item{\emph{Keep in mind that some scientists will not follow centralized guidelines and/or will pursue their own projects which may conflict with attempts to centralize collaboration-wide efforts.} The EPOC actively encouraged EPO participation amongst collaboration members both on initiatives specifically coordinated for DES and on individual projects. In some cases, we worked to consolidate personal and collaboration-sponsored efforts. This was often difficult, however, as individual members would not communicate their efforts with the EPOC. As it is impossible to change individual behavior and enforce such lines of communication, we strongly suggest that future EPO committees develop official policy (e.g., quality standards, promotion rules) related to any collaboration-related EPO products. We stress that we encourage this coordinated effort largely because some EPO work that comes out of unilateral activities can be superb and should be distributed to the (likely) larger audience garnered by collaboration-sponsored activities.}

\item {\emph{Clearly distribute roles and responsibilities amongst EPO coordinators.} For the majority of programs listed in this article, the EPOC coordinators were responsible for developing new ideas, recruiting participation, implementing activities, and activity evaluation. Although this freedom was instrumental in the evolution of the EPO program, it also meant that it was sometimes difficult to assess an activity objectively. We recommend assigning explicit roles or creating advisory positions to ensure that resources are being allocated practically and that project implementation aligns with project goals.}

\item{\emph{Consider organizing and/or attending science communication training sessions led by professionals.} Finally, we strongly recommend the organization of science communication workshops and training. We believe that DES collaboration members, including the coordinators of the EPOC, would significantly benefit from the professional experience of science communication experts. If scientists are expected to be at the forefront of knowledge, they should be cognizant of the most effective means to communicate that knowledge.}
\end{enumerate}

\section*{Acknowledgments}

The authors thank all of the DES members who contributed time and effort to the DES EPO program. R.~C.~Wolf is grateful to E.~Bechtol and S.~Pasero for useful discussions and professional EPO expertise.

Reference herein to any specific commercial product, process, or service by trade name, trademark, manufacturer, or otherwise, does not constitute or imply its endorsement by the United States Government or the Jet Propulsion Laboratory, California Institute of Technology.

\section*{Contributions to the article}
This section outlines specific contributions to this article.

\subsection*{Design, writing, and correspondence}
R.~C.~Wolf is responsible for the overall article design and the majority of the writing. Wolf provided analysis and is the corresponding author for this work.

\subsection*{Supporting writing and analysis}
A.~K.~Romer and B.~Nord provided insight regarding the organization and writing of specific sections, in particular for those related to projects in which they had unique knowledge or experience. As an EPO and communications professional, L.~Biron provided valuable perspective and feedback. K.~Bechtol and J.~Zuntz provided valuable feedback during the writing process as internal reviewers.

\subsection*{Data acquisition}
A.~Fert\'e gathered social media analytics data and J.~M.~Hislop obtained internal assessment data.

\section*{Contributions to Online DES EPO}
Here we provide a breakdown of major contributions to the online DES EPO initiatives, without which the work described in this paper would not be possible.

\subsection*{Social Media Management}
R.~C.~Wolf provided essential effort in maintaining consistent social media output for three years. 

\subsection*{Dark Energy Detectives}
Many DES Scientists contributed articles to this effort. B.~Nord edited and provided content, and originated the concept and design.

\subsection*{Web Site}
Kevin Munday, Carolina Manel, and Xenomedia were paid for their work to build the DES Web Site. B.~Nord led the design and content development. 

\subsection*{DArchives}
The DArchive originated when R.~Cawthon, B.~Nord, and R. Wolf sought a method to convey DES papers more directly to the public. Editing, writing, and organizing was performed by R.~Cawthon, R.~C.~Wolf, A.~Kremin, M.~S.~S. Gill, B.~Nord, and A.~K.~Romer. Many DES scientists contributed content and time. 

\subsection*{DEST4TD}
A.~K.~Romer and R.~C.~Wolf were the principals for designing and organizing this initiative to bring consistent connection between scientists and the public. T.~Lingard wrote code to automate scientist recruitment. J.~M.~Hislop subsequently developed and maintained the code. Many DES scientists contributed content.

\subsection*{DarkBites}
The content of this program was made possible by C.~Chang, D.~Q.~Nagasawa, R.~R.~Gupta, J.~Muir, and E.~Suchyta. B.~Nord originated the concept, based on an observations of how some science communicators successfully approach social media interactions. 

\subsection*{DEScientist of the Week}
R.~C.~Wolf initiated and carried out this effort to humanize scientists and normalize the science profession in broader culture.

\subsection*{Multilingual EPO}
The MLDES project would not have been possible without the many collaboration members who volunteered their time to translate various social media posts (including DEST4TD and DarkBites). S.~Avila, A.~ M{\"o}ller, A.~A.~Plazas, and I.~Sevilla-Noarbe were primarily responsible for translating content for the Spanish-language version of the website. T.~S.~Li and Y.~Zhang translated content and managed the DES Weibo account.

\subsection*{Image Curation}
S.~Hamilton and R.~Das lead the effort for DES image curation and management. This included DES science images, as well as other photographs, slides, and images that could be used for EPO activities.

\subsection*{YouTube}
The DES Youtube channel and other visual media have primarily been managed by C.~Krawiec.


\section*{Funding and Support}
AAP is supported by the Jet Propulsion Laboratory. Part of the research was carried out at the Jet Propulsion Laboratory, California Institute of Technology, under a contract with the National Aeronautics and Space Administration.

Parts of this research were conducted by the Australian Research Council
Centre of Excellence for All-sky Astrophysics (CAASTRO), through project
number CE110001020.

Funding for the DES Projects has been provided by the U.S. Department of Energy, the U.S. National Science Foundation, the Ministry of Science and Education of Spain, 
the Science and Technology Facilities Council of the United Kingdom, the Higher Education Funding Council for England, the National Center for Supercomputing 
Applications at the University of Illinois at Urbana-Champaign, the Kavli Institute of Cosmological Physics at the University of Chicago, 
the Center for Cosmology and Astro-Particle Physics at the Ohio State University,
the Mitchell Institute for Fundamental Physics and Astronomy at Texas A\&M University, Financiadora de Estudos e Projetos, 
Funda{\c c}{\~a}o Carlos Chagas Filho de Amparo {\`a} Pesquisa do Estado do Rio de Janeiro, Conselho Nacional de Desenvolvimento Cient{\'i}fico e Tecnol{\'o}gico and 
the Minist{\'e}rio da Ci{\^e}ncia, Tecnologia e Inova{\c c}{\~a}o, the Deutsche Forschungsgemeinschaft and the Collaborating Institutions in the Dark Energy Survey. 

The Collaborating Institutions are Argonne National Laboratory, the University of California at Santa Cruz, the University of Cambridge, Centro de Investigaciones Energ{\'e}ticas, 
Medioambientales y Tecnol{\'o}gicas-Madrid, the University of Chicago, University College London, the DES-Brazil Consortium, the University of Edinburgh, 
the Eidgen{\"o}ssische Technische Hochschule (ETH) Z{\"u}rich, 
Fermi National Accelerator Laboratory, the University of Illinois at Urbana-Champaign, the Institut de Ci{\`e}ncies de l'Espai (IEEC/CSIC), 
the Institut de F{\'i}sica d'Altes Energies, Lawrence Berkeley National Laboratory, the Ludwig-Maximilians Universit{\"a}t M{\"u}nchen and the associated Excellence Cluster Universe, 
the University of Michigan, the National Optical Astronomy Observatory, the University of Nottingham, The Ohio State University, the University of Pennsylvania, the University of Portsmouth, 
SLAC National Accelerator Laboratory, Stanford University, the University of Sussex, Texas A\&M University, and the OzDES Membership Consortium.

KR acknowledges support from the Science and Technologies Research Council, in particular from grant awards ST/N001087/1, ST/P000525/1, ST/M003574/1.

\bibliographystyle{apacite}
\bibliography{sample}

\appendix

\section{DES Project and Science}
\label{sec:sciproj}
In the late 1990s, two teams of astronomers made the unexpected discovery that the Universe is expanding at an accelerating rate \citep{R98,Perlmutter}. The mysterious agent of this acceleration, which acts against gravity's attractive force, has been named ``dark energy.'' Understanding the nature of dark energy has become one of the greatest unsolved problems in modern cosmology \citep{wmap,Planck}. The goal of the international DES collaboration is to study this accelerating expansion with unprecedented precision and accuracy.

DES is surveying 5000 deg$^2$ of the southern sky, using the Dark Energy Camera (DECam) \citep{Flaugher} mounted on the 4-$m$ Victor M. Blanco telescope at the Cerro Tololo Inter-American Observatory (CTIO) in Chile. DES is scheduled to take data for five years (2013-2018), observing each year from August-February. Although much of the observing is computer-automated, DES collaboration members travel to the telescope site during the DES season to help take data. Once DES data is collected, the DES Data Management team stores and processes the data, preparing it for DES scientists all over the world to analyze. DES traces its origins as a project concept back to at least 2004. However, the first DES images were not taken until September 2012. 

One of the unique strengths of DES is that it employs four complementary techniques to study the effects of dark energy, through observations of: Type Ia supernovae;  gravitational lensing; galaxy clusters; and baryon acoustic oscillations. During the course of the survey, DES will observe thousands of supernovae, map millions of galaxies, and measure the growth of large-scale structure of our universe \citep{DES}.

In addition to studying fundamental cosmological probes, DES makes important contributions to astronomy. DES scientists study the outer reaches of our solar system, finding new candidates for dwarf planets \citep{Gerdes} and other trans-Neptunian objects. They identify galactic neighbors to our Milky Way \citep{Li}. They search for optical counterparts to newly discovered gravitational waves \citep{GW}.

\begin{landscape}
\begin{figure}[htp]
\centering
\includegraphics[scale=0.8,angle=-90]{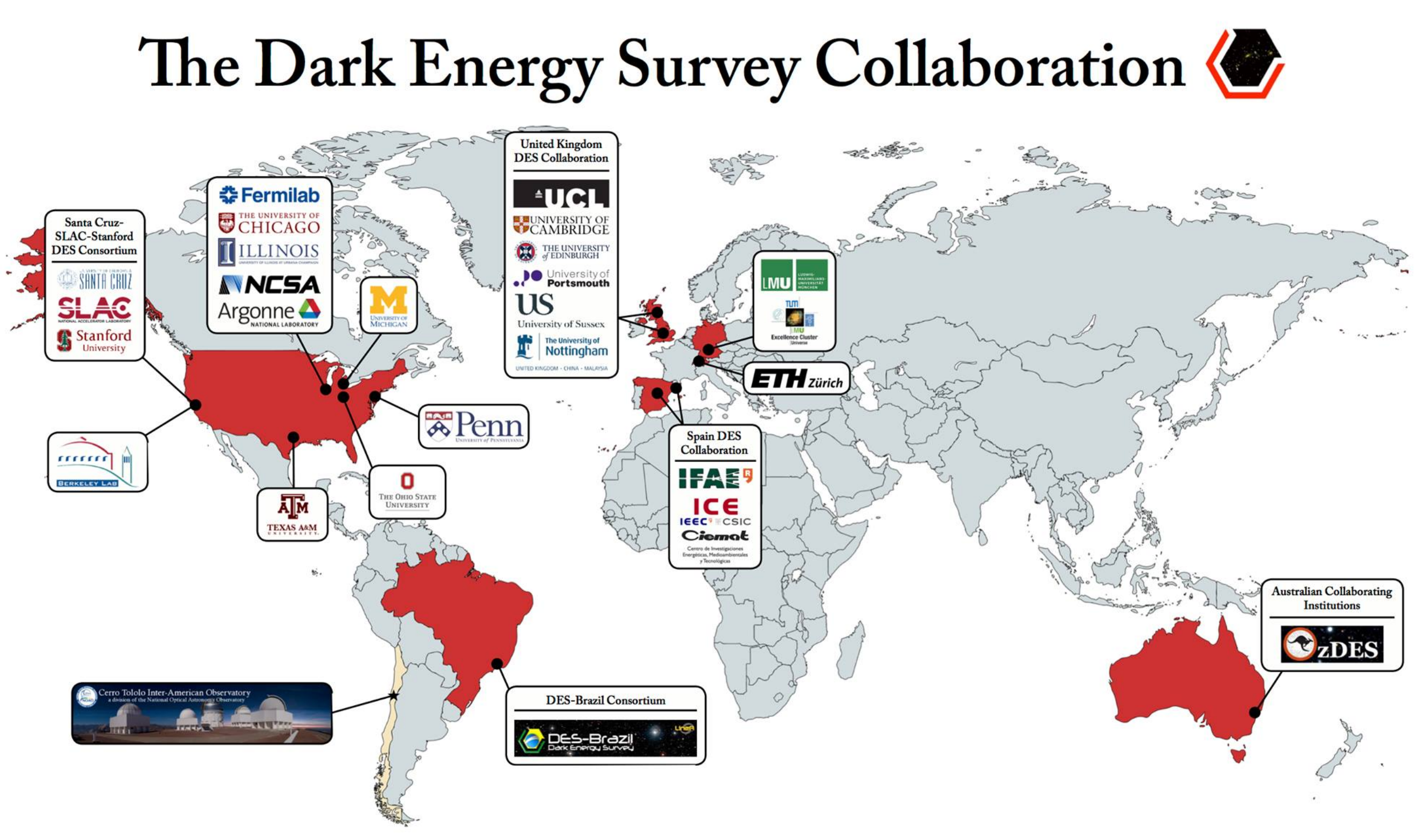}
\caption{Map of DES collaborating institutions. Figure credit: Judit Prat, IFAE, DES-Spain.}
\label{fig:desmap}
\end{figure}
\end{landscape}

DES is a collaboration of over 500 scientists from 25 institutions in seven countries around the world (a map of DES institutions is shown in Figure~\ref{fig:desmap}). University faculty and researchers, laboratory and observatory staff scientists, postdoctoral researchers, graduate students, and undergraduates are all working to answer unanswered questions about our Universe. The support staff, at the telescope and at DES institutions, enable DES scientists to travel for observing and to gather to discuss latest results at conferences and collaboration meetings. Together, members of the DES collaboration are at the cutting-edge of science and forging a new frontier for large-scale astronomy.

\begin{figure}
\centering
\includegraphics[scale=0.6]{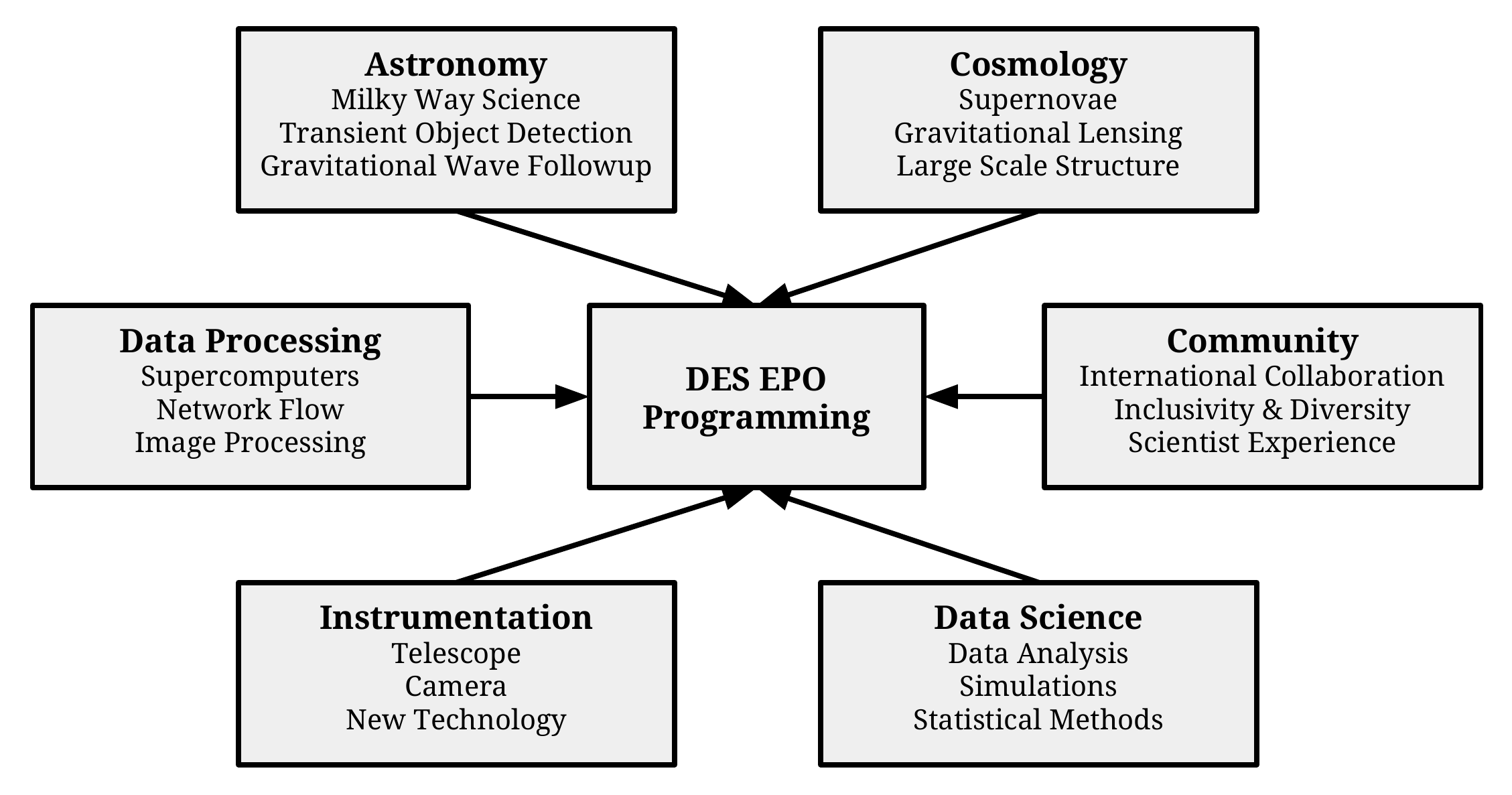}
\caption{Schematic diagram illustrating various components of DES which provide inspiration for the EPO effort. Examples included here are merely a subset.}
\label{fig:epochart}
\end{figure}

The various aspects of the survey highlighted in this section are summarized in Figure~\ref{fig:epochart}. The DES EPO program draws inspiration from each of these components to design innovative EPO programming without necessary relying on published data products. Examples included in the Figure represent only a subset of the material available for EPO programming.

\subsection{Organization and Management}
\label{subsec:org}

The three signatory institutions of DES are the Fermi National Accelerator Laboratory (hereafter Fermilab), National Center for Supercomputing Applications (hereafter NCSA), and the National Optical Astronomy Observatory (hereafter NOAO). Support for DES is provided by grants from these respective institutions, primarily from the U.S. Department of Energy and the National Science Foundation. Members of the DES \emph{Project Office} report directly to these agencies.

DES Scientists are categorized into \emph{members}, \emph{participants}, and \emph{external collaborators}.\footnote{DES membership policies and infrastructure tasks are described in: \texttt{http://www.darkenergysurvey.org/wp-content/uploads/2016/05/ \newline membership\_policy\_revised-Dec-2011.pdf}} DES members are senior scientists, including faculty (tenured and tenure-track) and senior research associates, at official DES collaborating institutions. Participants are typically current postdoctoral researchers and graduates students of DES members. Members and participants have access to DES data and data products. External collaborators are senior scientists at non-DES institutions who provide resources that are otherwise unavailable to the collaboration, e.g., access to private telescopes. Participants can gain permanent access to DES data by working on DES infrastructure. Infrastructure activities include work on DES instrumentation, pipeline development, data calibration, and management activities. After one year of Full Time Equivalent (FTE) infrastructure work, participants can apply for data rights; after 2 FTE, participants can apply for Builder status, which includes data rights and optional authorship on DES papers. Throughout this work, we will refer to our DES colleagues as ``collaboration members." This includes full DES members, participants, and external collaborators.

There are several channels available for DES scientists to communicate within working groups and across the collaboration. All scientists are encouraged to subscribe to a DES-wide LISTSERV (electronic mailing list) which is used for collaboration updates. Twice a month there are collaboration-wide telephone conferences where scientists can share results and receive feedback. Collaboration leadership also organizes two annual in-person week-long meetings where scientists congregate to discuss progress and results.

The internal organization of DES is divided into three main components: collaboration affairs, science, and operations. Collaboration affairs are overseen by the \emph{Management Committee}, who are responsible for making collaboration-wide decisions including membership and publication policy. The \emph{Science Committee} is responsible for managing the DES scientific program and ensuring all science requirements are met. Telescope operations, data management, and science releases are overseen by the \emph{Executive Committee}. Each of these three committees is further subdivided into a variety of smaller groups, e.g. the Science Committee is comprised of science working groups and the Management Committee oversees the Publications Board (who review DES publications and enforce DES publication policy) and Speakers' Bureau (who recruit DES members to speak at conferences on behalf of the collaboration). Each subcommittee is governed by official protocol that dictates how collaboration members should work both within the respective committee, and with the collaboration as a whole.\footnote{Further detail on DES policies and organization can be found in the DES Memorandum of Understanding: \texttt{https://www.darkenergysurvey.org/wp-content/uploads/2016/05/DES\_MOU\_as\_executed.pdf}}

\begin{figure}[htp]
\centering
\includegraphics[scale=0.4]{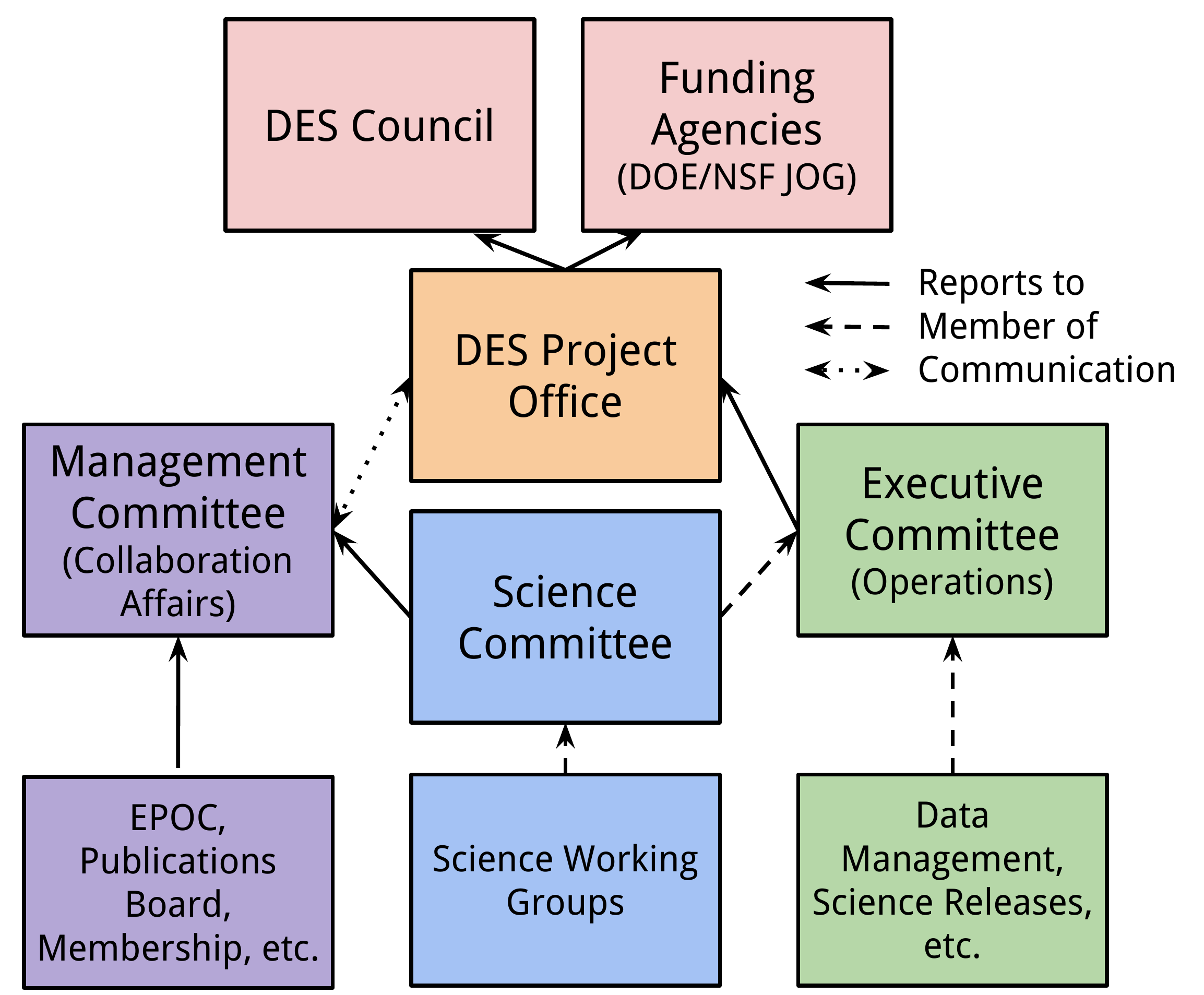} 
\caption{DES internal organization chart, including the EPOC (purple), adapted from the DES director's presentation at the Fall 2016 collaboration meeting. Solid	arrows indicate	a group	that reports to and/or is appointed by the box to which it points. Dashed arrows indicate that the people named	in that	group are members of	the	higher-level Committee	to	which	that box points	(e.g., science working group coordinators are members of the Science Committee, and	the	Science	Committee co-chairs	are members of	the	
Executive	Committee).	Dotted	two-way arrows indicates a	line of	mutual communication. }
\label{fig:orgchart}
\end{figure}

The DES Education \& Public Outreach Committee (EPOC) became a part of the official DES organizational structure in the Fall of 2014 and was placed under the umbrella of collaboration affairs (for details regarding the creation and development of the EPO Committee, see Appendix~\ref{subsec:epoevol}). Prior to that time, DES did not have a centralized EPO effort nor official recognition of EPO on a collaboration-wide scale. As such, once the EPOC formed, there were no policies in place for how the EPOC and its programming should interact and coordinate with the rest of the collaboration. For example, representatives from the EPOC are not invited to Management Committee meetings, although other committees responsible for collaboration affairs are included. A summary of the current DES organizational structure, including the EPOC, is presented in Figure~\ref{fig:orgchart}.

The roles and responsibilities of the EPOC have evolved organically since its inception. As the primary organizers of EPO for the collaboration, the EPOC oversees and contributes to: updating and maintenance of the DES website, DES social media, creation of informal and formal educational materials, DES events with local communities (e.g., museum events and science fairs), internal EPO reporting, public relations,\footnote{Note that this is distinct from official press releases which are organized through the Fermilab press office.} and much more. The centralized DES EPO program has a limited, floating budget per the discretion of the DES director, which is jointly funded by the DES collaborating institutions. Details of how these funds have been allocated thus far are discussed in Section~\ref{sec:programs}.


\subsection{The Evolution of EPOC}
\label{subsec:epoevol}

Prior to the Fall 2014 collaboration meeting at the University of Sussex, no sessions dedicated to EPO had been scheduled by the scientific organizing committee (SOC). Kathy Romer (a faculty member at Sussex) was the chair of the Sussex SOC and decided to arrange two EPO sessions. This was done in consultation with Brian Nord (a postdoctoral researcher at Fermilab), who had been, by then, running a DES EPO initiative known as \emph{The Dark Energy Detectives} blog (Section~\ref{subsubsec:DED}). Nord was not able to attend as he was observing in Chile, but recommended that Romer speak to Rachel Wolf (at the time, a graduate student at the University of Pennsylvania) about enhancing DES's social media presence (Romer's EPO focus to that date had been on working with school groups). And so began a grass-root effort to inspire coordinated EPO initiatives within DES. By the end of the Sussex collaboration meeting, Romer and Nord were asked by the DES director to lead an official EPO committee (the EPOC) for DES. They agreed to do so on the condition that Wolf was also included. Nord, Romer and Wolf (NRW hereafter) thus officially became the coordinators of the EPOC. 

To discuss the organization and implementation necessary to get DES EPO off the ground, NRW established weekly (EPOC coordinator) telecons. NRW also created an internal DES EPO LISTSERV to communicate about EPO projects and opportunities with colleagues.



During the first year of the EPOC, most programming was organized and executed by NRW. Much of the effort was focused on updating, enhancing and maintaining the DES online presence. At the following semi-annual collaboration meeting (Spring 2015), NRW organized several EPO-specific sessions to present the work carried out so far, to receive feedback, and to generate new ideas. There were no shortage of new ideas and it became clear that more colleagues would need to be recruited to keep up with demand. Fortunately several DES members were eager to contribute, and even to lead, certain EPO projects, especially those that appealed to their particular interests (e.g., writing, artwork, or astrophotography). In addition to the weekly NRW coordinator meetings, monthly EPOC telecons were established to discuss the progress of the various projects. One of those projects is internal communication and resulted in a monthly DES-EPO newsletter that is sent electronically to every registered scientist in the DES membership database. 



\section{EPO Program Logic Models}
\label{sec:lm}

In this Appendix we present logic models for The Darchive, DEST4TD, DarkBites, and DEScientist of the Week EPO Initiatives. In each model, we describe the inputs, actions, and product outputs created by the EPO coordinators and participating scientists. We also present the DES-specific outcomes, as well as the predicted short-term, medium-term, and long-term outcomes. We encourage readers to treat these logic models as planning outlines for the respective projects and leave details on project evaluation for future work.

\begin{figure}[htp]
\includegraphics[scale=0.55]{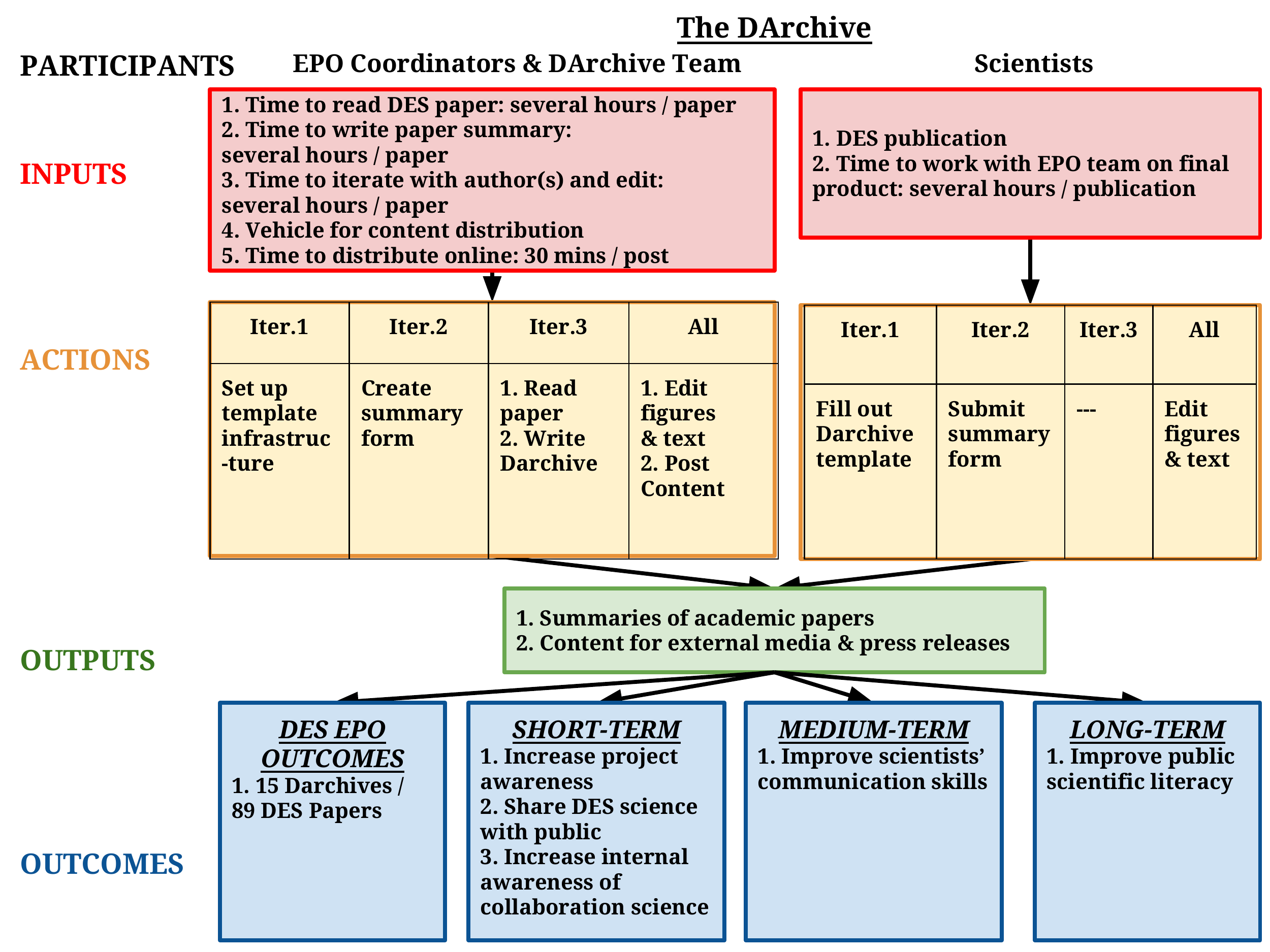}
\caption{Logic model describing The DArchive project structure and outcomes.}
\label{fig:darchivelm}
\end{figure}

\begin{figure}[htp]
\includegraphics[scale=0.55]{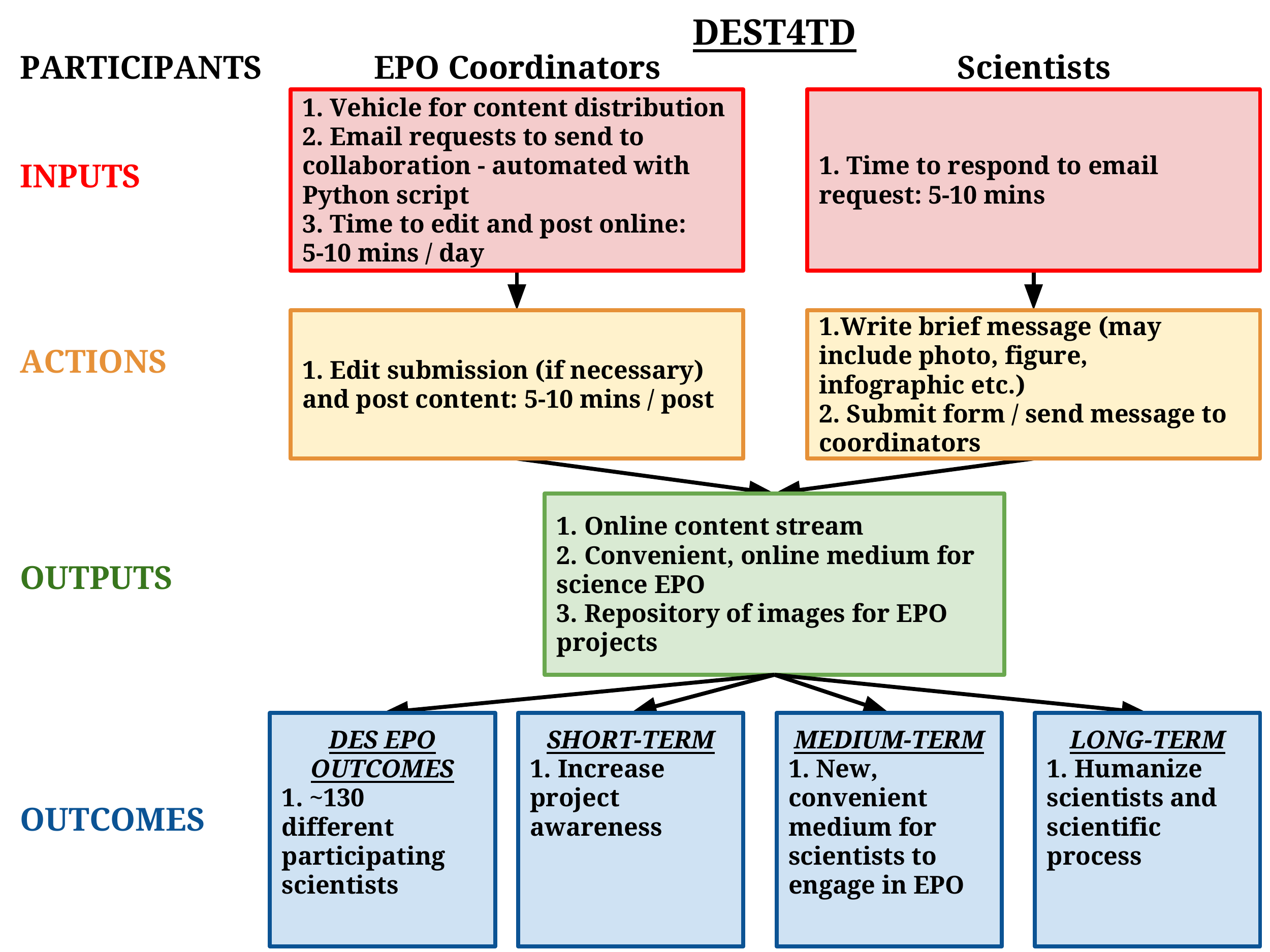}
\caption{Logic model describing DEST4TD project structure and outcomes.}
\label{fig:dest4tdlm}
\end{figure}

\begin{figure}[htp]
\includegraphics[scale=0.55]{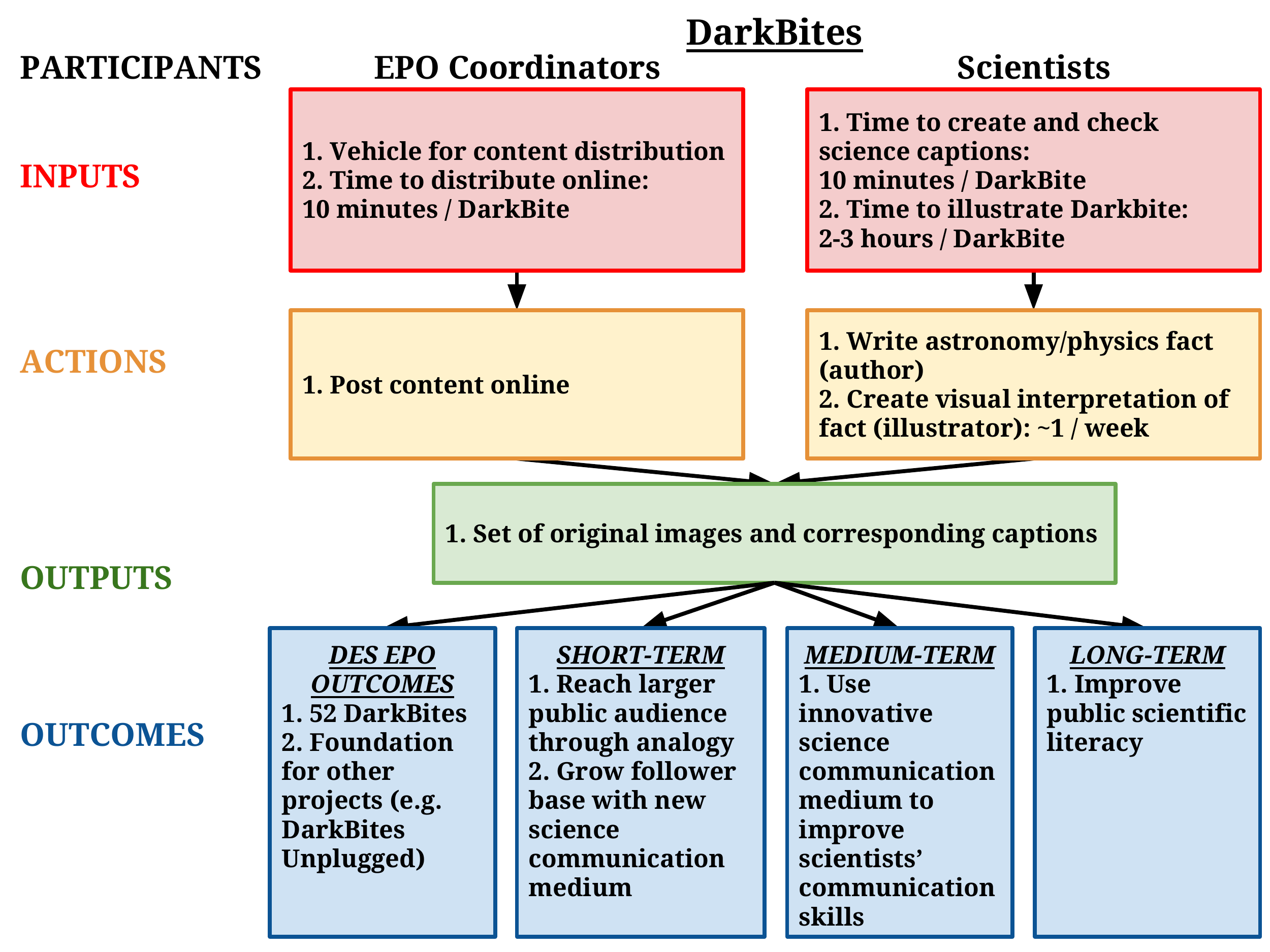}
\caption{Logic model describing DarkBites project structure and outcomes.}
\label{fig:darkbiteslm}
\end{figure}

\begin{figure}[htp]
\includegraphics[scale=0.55]{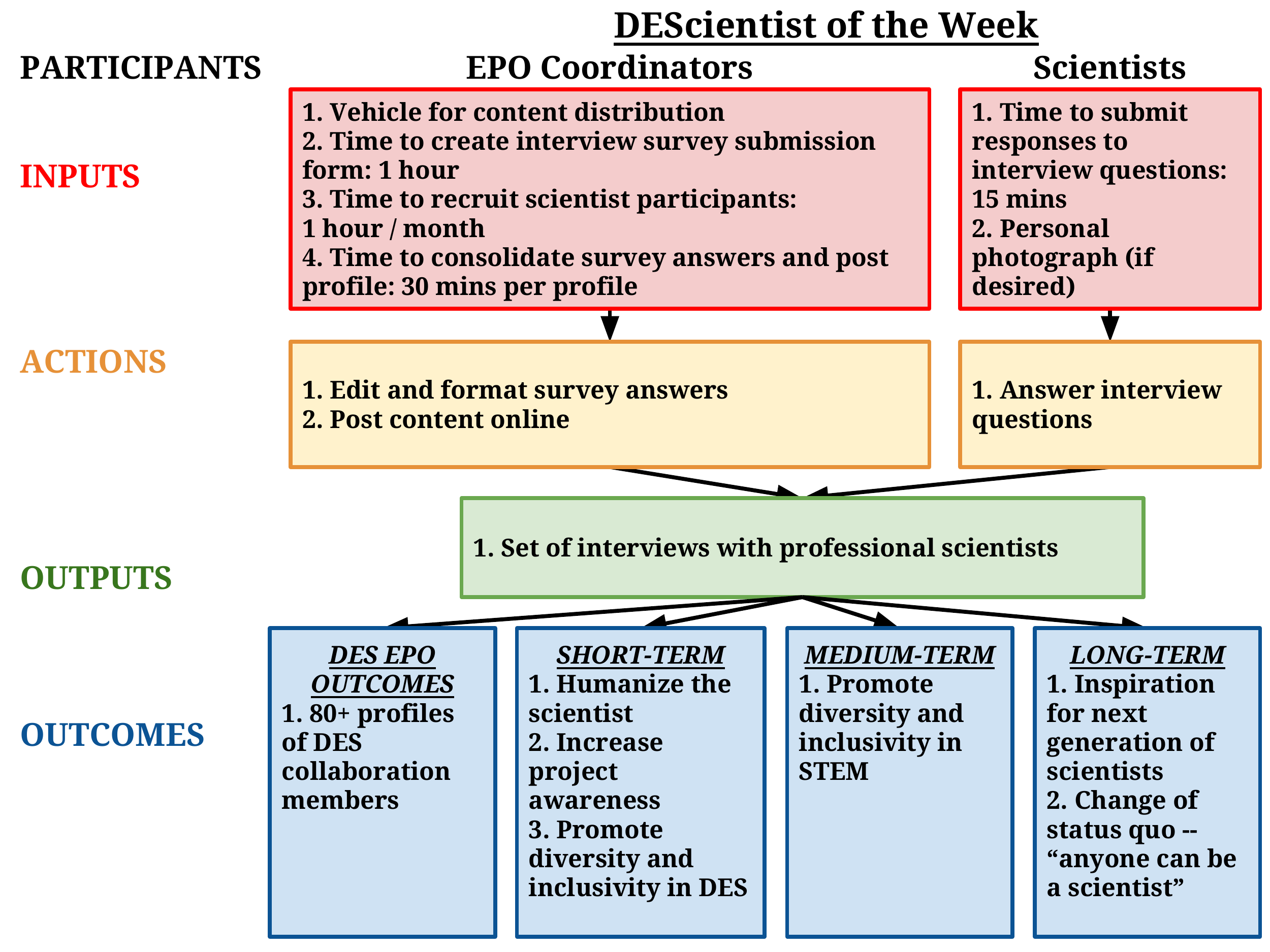}
\caption{Logic model describing DEScientist of the Week project structure and outcomes.}
\label{fig:desowlm}
\end{figure}

\clearpage

\section{Internal EPO Survey}
\label{sec:intersur}

Approximately one year after the formation of the EPOC, we conducted a survey to gauge collaboration members' awareness of and attitudes towards the DES EPO program in general as well as four specific EPO initiatives. We created an online survey and emailed it to the entire collaboration, asking all collaboration members to participate. We also advertised the survey at the Fall 2015 collaboration meeting. DES stickers were offered to those who participated in the survey. A total of 90 collaboration members participated. 

Here we include a transcribed copy of the survey. Asterisks indicate questions which required a response. Participants were offered the opportunity to exit the survey after a few key sections; these are indicated with investigator notes. The online version of the survey included examples of the particular EPO initiatives for those who indicated they were unfamiliar with a particular project. We do not include these examples here.



\section*{Supplementary material}

# DES Education & Public Outreach (EPO) Feedback Form

Thanks for checking out the EPO evaluation form! We'd very much like to know your thoughts about DES EPO projects.

The form is divided into several sections. There are four sections about specific outreach projects and some general questions at the end. The questionnaire is a bit long (filling out the whole form may take 20 minutes), but you don't have to complete the entire thing!

Thanks for your participation!
The EPO Committee
----------------------------

**\*Name:**

**\*Institution:**

**\*Email Address:**

**\*Current Position**
☐ Graduate Student ☐ Postdoc ☐ Faculty/Professor ☐ Staff Scientist ☐ Other:

## DES SOCIAL MEDIA (SM) PROJECTS

**\*What social media platforms do you use?**
Please select any that apply
☐ Facebook ☐ Twitter ☐ Instagram ☐ Snapchat ☐ None ☐ Other:

**\*Do you know where to find DES social media posts?**
☐ Yes ☐ No ☐ What's social media??

**DES Thought for the Day (DEST4TD)**
**\*Are you familiar with the DES Thought for the Day (DEST4TD) project?**
☐ Yes ☐ No ☐ Yes, but I could use a refresher
*\*\*Investigator Note: If the participant indicated anything other than "Yes", he/she was sent to an example DEST4TD.\*\**

**Check all that apply**
☐ I have contributed a DEST4TD ☐ I would contribute a DEST4TD
☐ I think DEST4TD is a worthwhile EPO initiative ☐ I think DEST4TD can be improved

**T4TD Feedback**
Looking for inspiration - What do you think of the T4TD posts? Have you been asked to submit a T4TD? What would encourage you to participate in T4TD?

**DEScientist of the Week**
**\*Are you familiar with the DEScientist of the Week project?**
☐ Yes ☐ No ☐ Yes, but I could use a refresher
*\*\*Investigator Note: If the participant indicated anything other than "Yes", he/she was sent to an example DEScientist of the Week.\*\**

**Check all that apply**
☐ I have participated in DEScientist of the Week ☐ I would participate in DEScientist of the Week
☐ I think DEScientist of the Week is a worthwhile EPO project ☐ I think DEScientist of the Week can be improved

**DEScientist of the Week (SoW) Feedback**

Looking for inspiration - What do you think of SoW?  Would you be willing to participate in SoW?
_____

**Would you like to continue to questions about The DArchive?**
☐ Sure!    ☐ No thanks, I'd like to go to the final page.
*\*\*Investigator Note: As the DArchive is a longer project, we gave participants the opportunity to end the survey at this point.\*\**

**The DArchive**
**Are you familiar with The DArchive project?**
☐ Yes, and I've read a post recently    ☐ Yes, but I haven't read a post before
☐ No                                     ☐ Yes, but I could use a refresher
*\*\*Investigator Note: If the participant indicated anything other than "Yes, and I've read a post recently", he/she was sent to an example DArchive Post.\*\**

**Check all that apply**
☐ I have submitted a DArchive summary          ☐ I would be willing to write a DArchive summary
☐ I think the DArchive is a worthwhile EPO project    ☐ I think the DArchive can be improved

**The DArchive Feedback**
Looking for inspiration - What do you think of The DArchive? Do you think the posted DArchives summarized the DES papers well? Would you be willing to write a DArchive summary?
_____

**Would you like to continue to questions about Dark Energy Detectives?**
☐ Sure!    ☐ No thanks, I'd like to go to the final page.
*\*\*Investigator Note: As Dark Energy Detectives is a longer project, we gave participants the opportunity to end the survey at this point.\*\**

**Dark Energy Detectives (DED)**
**Are you familiar with The Dark Energy Detectives project?**
☐ Yes    ☐ No    ☐ Yes, but I could use a refresher
*\*\*Investigator Note: If the participant indicated anything other than "Yes", he/she was sent to an example Dark Energy Detectives.\*\**

**Check all that apply**
☐ I have written a DED post                    ☐ I would be willing to write a DED post
☐ I think the DED is a worthwhile DES project  ☐ I think DED can be improved

**Dark Energy Detectives Feedback**
Looking for inspiration - What would encourage you to write a post?  Do you like the theme of the blog?
_____

**Last Few General Questions**
You're almost done!

**\*Have you received an email to participate in a DES EPO project?  Did you participate?  If not, what would inspire you to get involved?** You can answer for one project in particular or several.
_____

**\*Do you read the monthly EPO Newsletter?  Do you find it useful?  Do you think it can be improved?**
_____

**\*The EPO Committee is working to maintain a record of all DES EPO activity.  Did you know you can record non-DES STEM outreach activity?**
☐ Yes, and I have submitted a form    ☐ Yes, but I have not submitted the form
☐ No                                   ☐ No, I thought it was for only DES-specific events



\end{document}